\documentclass[journal=jctcce,manuscript=article,layout=onecolumn]{achemso}

\usepackage[version=3]{mhchem} 
\usepackage{booktabs}
\usepackage{xcolor}
\usepackage{amsmath}

\author{Giovanni Pireddu}
\affiliation{Dipartimento di Chimica e Farmacia, Universit\`a degli Studi di Sassari, Via Vienna 2, 01700 Sassari, Italy}
\alsoaffiliation{Institute of Physics, Polish Academy of Sciences, Al. Lotnik\'ow 32/46, 02-668 Warsaw, Poland}
\email{gpireddu@uniss.it}
\author{Federico G. Pazzona}
\affiliation{Dipartimento di Chimica e Farmacia, Universit\`a degli Studi di Sassari, Via Vienna 2, 01700 Sassari, Italy}
\author{Pierfranco Demontis}
\affiliation{Dipartimento di Chimica e Farmacia, Universit\`a degli Studi di Sassari, Via Vienna 2, 01700 Sassari, Italy}
\author{Magdalena A. Za\l{}uska-Kotur}
\affiliation{Institute of Physics, Polish Academy of Sciences, Al. Lotnik\'ow 32/46, 02-668 Warsaw, Poland}

\title{Scaling-up Simulations of Diffusion in Microporous Materials}

\begin{document}

\begin{abstract}
We introduce and demonstrate the coarse-graining of static and dynamical properties of host-guest systems constituted by methane in two different microporous materials. 
The reference systems are mapped to occupancy-based pore-scale lattice models. 
Each coarse-grained model is equipped with an appropriate coarse-grained potential and a local dynamical operator, which represents the probability of inter-pore molecular jumps between different cages.
Both the coarse-grained thermodynamics and dynamics are defined based on small-scale atomistic simulations of the reference systems.
We considered two host materials: the widely-studied ITQ-29 zeolite and the LTA-zeolite-templated carbon, which was recently theorized. 
Our method allows representing with satisfactory accuracy and a considerably reduced computational effort the reference systems while providing new interesting physical insights in terms of static and diffusive properties.
\end{abstract}

\section{Introduction}

Computer simulations of physical systems have widely demonstrated their usefulness in understanding complex phenomena, by both offering a direct comparison with purely theoretical approaches, and being capable of providing insightful predictions.\cite{Allen2017,Frenkel2002,Parrinello1985}

Over the last three decades, multiscale modelling approaches have progressively gained interest in several disciplines and for different applications.\cite{Levitt1975,Chen1998}
In particular, bottom-up protocols allow for representing the systems of interest in increasingly large time- and length-scales, by progressively decreasing the level of detail associated with each representation through coarse-graining methods.\cite{Bilionis2013,Izvekov2005,Izvekov2005b,Noid2008,Reith2003,Tsourtis2017,Ma1976}
However, there is still a lack of a general method for mapping a fine-grained (FG) representation to a coarse-grained (CG) (i.e.~less-detailed) one, and the choice of such transformation is usually system-dependent. 

In this work, we focus on the mesoscopic representation of host-guest systems constituted by microporous materials and gas molecules. 
Nowadays, microporous materials are broadly employed in different scenarios and for different scopes, such as gas storage, separation of mixtures, heterogeneous catalysis, etc.\cite{ZeoHandbook2003,Furukawa2013}
Many of the processes involved in such applications strongly depend on the adsorption and diffusion behaviour of the guest molecules in the porous environment.\cite{Karger1992}
Thus, a general and sufficiently accurate mesoscale modelling framework for such phenomena could help to explain diffusive and sorptive properties and allow testing new systems \emph{in-silico}, such as hypothetical sorbent materials for various applications.\cite{Deem2009}

Lattice models of host-guest systems have demonstrated the capability of representing adsorption and diffusion phenomena with a considerably smaller computational effort compared to atomistic methods and yet allowing to reproduce the properties of interest with satisfactory accuracy.\cite{Snurr1994,Saravanan1998,Toktarbaiuly2018,Pazzona2009,Pazzona2009b}

In our case, we map the reference systems into pore-scale lattice models, in which each node represents a pore or a cage of the host material and is equipped with an occupancy state $n$ indicating the number of guest molecules present in such pore of the reference material.
Both the thermodynamics and the mass-transfer dynamics in the CG representations are modelled to match with the results of FG atomistic simulations.
In particular, we performed small-scale grand-canonical Monte Carlo (GCMC) simulations to retrieve the static properties, and relatively short canonical molecular dynamics (MD) simulations to model the transition-rates associated with the inter-cage jumps performed by the guest molecules.

To demonstrate the capabilities of our method, we chose to represent two interesting systems constituted by two different host materials and involving methane molecules as the guest species. 
The first material is the all-silica ITQ-29 zeolite, which is a well-studied material for the modelling of cage-to-cage dynamics and diffusion of small molecules in microporous materials.\cite{Dubbeldam2005,Tunca2002}
The second material is the LTA-zeolite-templated carbon (which we will refer to as ZTC), recently introduced as hypothetically obtainable by carbon-templating the afore-mentioned zeolite.\cite{Braun2018}

Zeolite-templated carbons are a relatively new class of nanoporous carbon materials, which exhibit peculiar properties when employed as methane sorbents.\cite{Stadie2015,Stadie2013}
Despite being related to its zeolite precursor and having the same topology in terms of pore connectivity, the ZTC we consider is structurally different as it presents larger free-volume in each pore and significantly larger openings between neighboring cages. 
For this reason, it is particularly interesting to compare the static and dynamic properties of the two systems.

The remainder of this work is organized as follows: in the section Methods, we introduce our CG method and explain our experimental setup for the numerical simulations; in the section Results and discussion, we show the numerical results in terms of comparisons between the two systems in terms of both static and dynamical properties; finally, in the last section we draw our conclusions, by highlighting the benefits and the limits of our method, and by proposing possible applications.

\section{Methods}\label{Sec:Methods}

\subsection{Coarse-grained structure}
\begin{figure}
  \includegraphics[width=3.25 in]{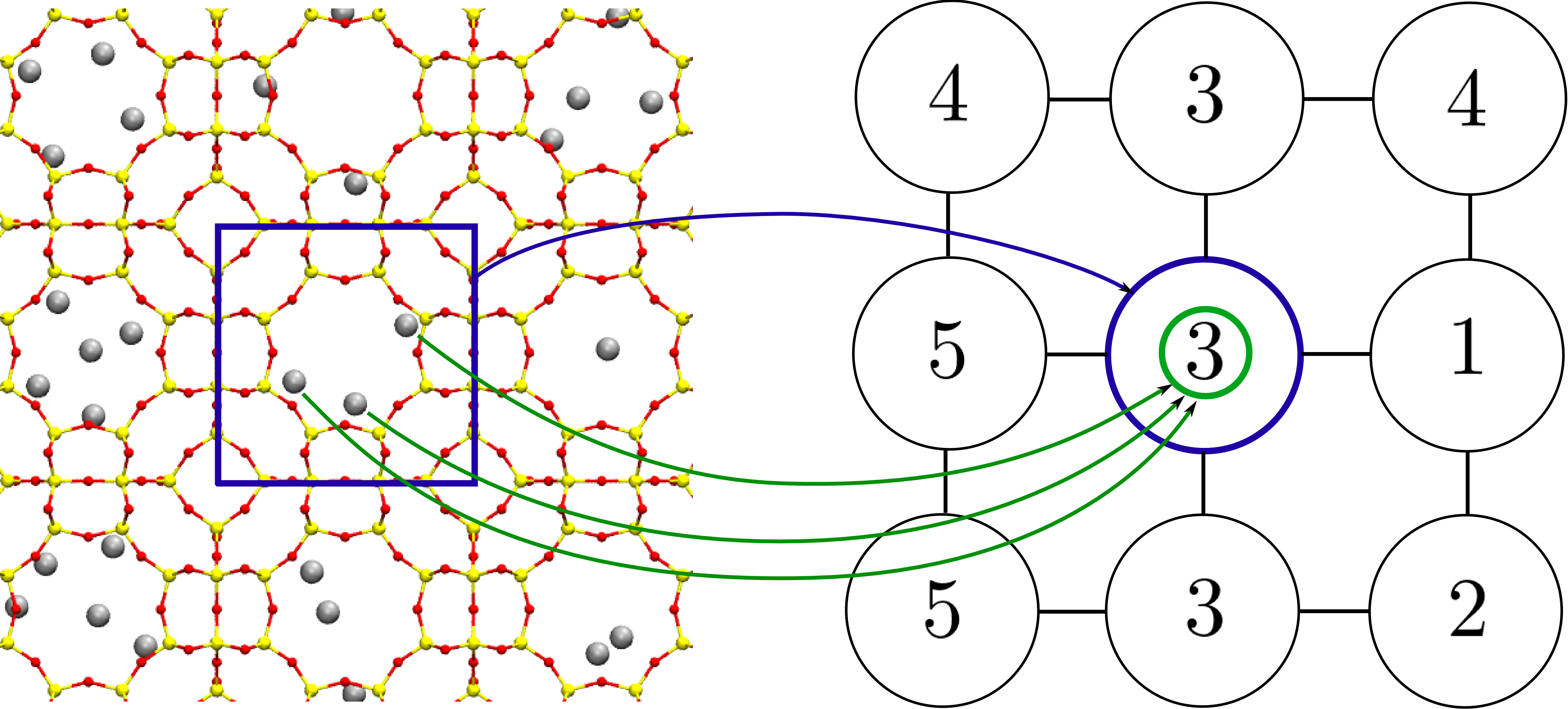}
  \caption{\footnotesize{Mapping of the methane/ITQ-29 system into its corresponding occupancy-based lattice model. The methane molecules are represented as grey spheres, while the framework atoms are represented by red (\ce{O} species) and yellow (\ce{Si} species) spheres. The blue color represents the mapping of the cavities of the host material, the green color indicates the mapping of the guest molecules. The links in the lattice model represent the connections between neighboring cages of the host material.}}
  \label{fgr:MAP}
\end{figure}

Our coarse-graining procedure begins with the structural definition of the lattice models, which represent the reference host-guest systems. 
Fig.~\ref{fgr:MAP} depicts a representation of the structural mapping of the molecular systems, from the atomistic picture to the occupancy-based lattice model.

We ideally tessellate the host materials with identical, non-overlapping cubic subvolumes called \textit{cells}. 
In our picture of the systems, each cell embeds a single pore of the reference host material. 
Since both the ZTC and the ITQ-29 present a simple cubic pore connectivity, the reference structures are conveniently mapped to cubic networks.
Each $i$-th cell of the CG lattice is then associated with its occupancy $n_i$, which corresponds to the total number of molecules whose center of mass falls within the $i$-th pore. 
In this fashion, the configuration of our lattice models is completely defined as the occupancy configuration $\mathbf{n}=\lbrace n_1, \dots, n_M \rbrace$, where $M$ is the total number of cells.

\subsection{Thermodynamics}
Following the Interacting-Pair-Approximation (IPA) approach,\cite{Pazzona2018} we associate the occupancy configurations of the lattice models with a CG potential function $\Omega$, which in the grand-canonical ensemble reads

\begin{equation}
    \Omega_\mu(\mathbf{n})= 
    \sum_i \left(H_{n_i} - \mu n_i\right)
    + \sum_{\langle ij \rangle} K_{n_i,n_j} , 
    \label{eqn:Whole_Sys_CGPot}
\end{equation}
where $\mu$ is the chemical potential, $H_{n_i}$ is the single-cell free-energy contribution of a cell with occupancy $n_i$, $K_{n_i,n_j}$ represents the free-energy contribution of the mutual interactions between neighboring cells with occupancies $n_i$ and $n_j$, and $\langle ij \rangle$ indicates a summation over nearest-neighboring cells.
For our purposes, we will assume that a parameterization of the interactions among the molecules residing in every single cell (through the $H_{n_i}$ parameter), along with a parameterization of the interactions among molecules located in neighboring cells (through the $K_{n_i,n_j}$ parameter) will suffice to adequately represent the effective interactions at a CG level of representation.
As shown in Eq.~\eqref{eqn:Whole_Sys_CGPot}, except from $\mu$, the CG potential function makes only use of local, occupancy dependent free-energy contributions.
The free-energy parameters are related to the respective partition function contributions via the following relations:
\begin{align}
 Q_n = e^{-\beta H_{n_i}}, \qquad Z_{n_i,n_j} = e^{-\beta K_{n_i,n_j}}.
\end{align}

Such contributions can be conveniently estimated using the following recurrence relations
\begin{equation}
    \frac{Q_n}{Q_{n'}}= 
    \frac{
      e^{-\beta\mu n}\, p^o_\mu(n)
    }{
      e^{-\beta\mu n'}\, p^o_\mu(n')
    },
    \label{eqn:Recur_Q}
\end{equation}
\begin{align}
  \notag  \frac{Z_{n_1,n_2}}{Z_{n'_1,n'_2}}= \left(\frac{e^{\beta \mu n'_1}Q_{n'_1}\,e^{\beta \mu n'_2}Q_{n'_2}}{e^{\beta \mu n_1}Q_{n_1}\,e^{\beta \mu n_2}Q_{n_2}}
\right)^{\frac{1}{\nu}} \\
\times \left( \frac{p_\mu(n'_1)\,p_\mu(n'_2)}{p_\mu(n_1)\,p_\mu(n_2)} \right)^{1 - \frac{1}{\nu}} \frac{p_\mu(n_1,n_2)}{p_\mu(n'_1,n'_2)} ,
    \label{eqn:recur_Z}
\end{align}
where $p^o_\mu(n)$, $p_\mu(n)$ and $p_\mu(n_1,n_2)$ are respectively the univariate occupancy distribution for a single-cell closed system, the univariate occupancy distribution of a single cell inside the reference system, and the bivariate occupancy distribution of two connected cells inside the reference system.
The symbol $\nu$ indicates the cell connectivity, which is $6$ for a 3D cubic network.

We estimate the local occupancy distributions from atomistic grand-canonical Monte Carlo (GCMC) simulations of the reference system.
Finally, the full set of free-energy contributions can be obtained by solving the recurrence relations in Eqs. \eqref{eqn:Recur_Q} and \eqref{eqn:recur_Z}, and using $H_0= 0$ kJ/mol and $K_{0,n}= 0$ kJ/mol (for every possible value of occupancy $n$) as starting points, since empty cells do not contribute to the total free-energy of the system.

\subsection{Elementary events for diffusion}

We assume that gas diffusion in microporous materials can be treated as the composition of several elementary and strictly local events. 
An appropriate observation time scale $\tau$ could allows to distinguish among single migration events occurring during the dynamical evolution of the host-guest systems. 
The aim of our work is to provide a stochastic modelling protocol to understand and represent such events, which we identify as single molecule jumps between two connected pores.
If the separability between elementary events holds, we assume that local dynamics can be represented by a local operator $W(\mathbf{m}'\mid\mathbf{m})$, which is applied to a single pair of connected cells and represents the transition probability of the transformation $(\mathbf{m}'\mapsto\mathbf{m})$ within the time interval $\tau$, where $\mathbf{m}=(n_1,n_2)$ and $\mathbf{m}'=(n'_1,n'_2)$ are the pair occupancy configurations before and after the transition.
For example, $W((4,6) \mid (5,5))$ represents the probability of a single molecule jump (from a cell of occupancy 5 to a neighboring cell of occupancy 5) resulting in the transformation $((5,5)\mapsto(4,6))$.
By following this approach, during each elementary event the local total mass $M_{12}=n_1 + n_2$ is conserved.

We empirically estimate the transition rate values $W$ from atomistic Molecular Dynamics (MD) simulations of the reference systems, during which we saved the positional configuration of the diffusing molecules in the system (all coordinates of methane molecules) every $\tau$ seconds, thus resulting in one trajectory of positional configurations for every MD simulation.
All trajectories of positional configurations obtained for the reference system are used to compute the time series of occupancy configurations; finally, the occupancies of each pair of connected pores between two consecutive occupancy configurations, say $(n_i(t),n_j(t))$ and $(n_i(t+\tau),n_j(t+\tau))$, are compared, and if the transformation from one occupancy pair to the other conserves mass [that is, if $n_i(t)+n_j(t)=n_i(t+\tau)+n_j(t+\tau)$], they are cumulated into the respective entries of $W$.

With this procedure, we obtain empirical values for each $W(\mathbf{m}'\mid\mathbf{m})$, where we ignore more complicated, multi-cell mass transfer mechanisms which may occur within the chosen time step, but still are much rarer than single jump events.

Since the migration of molecules from a pore to another is a thermally activated process, a common way of modelling jump rates is by introducing a temperature-dependent function of the free-energy barrier multiplied by a kinetic prefactor.
The first part is a static property, which accounts for the local free-energy change associated to an inter-cell jump event, whereas the kinetic prefactor $k_{M_{12}}$ models the frequency of the jump attempts frequency, and is a function of local occupancies.
We model the prefactor as a function of the local occupancy summation $M_{12}$, which is conserved during each elementary event.
The functional form we propose for the jump rates is the following:
\begin{align}
    W(\mathbf{m}'\mid\mathbf{m})= k_{M_{12}}e^{-\frac{\beta}{2} \left[\Omega_\mu({\mathbf{m}'})-\Omega_\mu({\mathbf{m})}\right]},
    \label{eqn:JRmodel}
\end{align}
with $\beta= 1/k_B T$, where $k_B$ is the Boltzmann constant and $T$ is the temperature.
The factor $1/2$ in the exponent on the right hand side of Eq.~\ref{eqn:JRmodel} stems from the detailed-balance (DB) condition imposed to a \emph{closed} pair of cells transforming from occupancy pair $\mathbf{m}$ to occupancy pair $\mathbf{m}'$, i.e.~$p_\mu(\mathbf{m})\,W(\mathbf{m}'|\mathbf{m})=p_\mu(\mathbf{m}')\,W(\mathbf{m}|\mathbf{m}')$, where $p_\mu(\mathbf{m})\propto\exp\{-\beta\Omega_\mu(\mathbf{m})\}$ and $k_{M_{12}}$ is symmetric with respect to the jump direction.
By following the definition of CG potential function given in Eq.~\ref{eqn:Whole_Sys_CGPot}, the potential function for a pair of connected cells, say cell $1$ and cell $2$, respectively occupied by $n_1$ and $n_2$ guest molecules, reads $\Omega_\mu(n_1,n_2)=-\mu (n_1+n_2) + H_{n_1} + H_{n_2} + K_{n_1,n_2}$.
By taking into account the fact that all the local transitions we consider do conserve mass, and by omitting the interaction contributions with the environment around the chosen pair of connected cells, the local change in free-energy due to the transition $\mathbf{m}=(n_1,n_2)\mapsto\mathbf{m}'=(n_1',n_2')$ is
\begin{align}\label{eqn:DeltaPot}
    & \Omega_\mu({\mathbf{m}'})-\Omega_\mu({\mathbf{m})}= H_{n'_1} + H_{n'_2} + K_{n'_1,n'_2} \\
    \notag & - \left(H_{n_1} + H_{n_2} + K_{n_1,n_2}\right).
\end{align}
Although the expression we proposed for the transition rates, $W(\mathbf{m}'\mid\mathbf{m})$ (see Eq.~\ref{eqn:JRmodel}), stems from the DB condition imposed on a closed pair of neighboring cells, the choice to not include the interactions with the neighbors around each pair hinders our operator from strictly fulfilling the DB condition on the whole system; however, this is consistent with our sampling scheme from the MD simulations, since we sample the transitions on the basis of each pair configuration only.
Of course, if we wanted to ensure that DB is strictly obeyed, the jump rates $W$ should also include information about occupancies in all the cells in the neighborhood around each pair; in other words, all such occupancies should appear as additional arguments in the conditionality of $W$. However, this would cost us a much heavier computational effort, that is required in order sufficiently robust statistics---this is against the spirit of our work, since we want to demonstrate how to coarse-grain molecular systems from relatively \emph{short} and \emph{small-scale} atomistic simulations. 
We also remark that modelling the full system by sampling a $W$ based merely on local pair occupancy configurations is equivalent to implicitly assume a mean-density around each pair, since the cells are embedded in the full system.
This method for the local dynamical evolution is analogue to a pair-wise stochastic evolution rule in a block cellular automaton, where we identify each block as a closed pair of connected cells.\cite{Toffoli1990}
We empirically found that our approximate model still yields a semi-quantitative matching of static properties in terms of occupancy histograms between the CG and MD simulations.

\subsection{Dynamical correlations}

If correlations between any two consecutive pore-to-pore jumps in the reference FG systems were negligible, then the reference systems could already be simulated by directly using the $W$ operators for the dynamical evolution of the lattice models with a Markov chain scheme. 
However, in real systems dynamical time-correlations, also called memory effects, may occur and significantly influence the diffusivity \cite{Nissila2002}. 
In principle, a higher-order (or higher-memory) model of the dynamics could be devised in such a way as to explicitly embed memory effects, and thus yield a realistic representation of the diffusion behaviour; however, also in this case, the amount of data that would be necessary for us to base such more sophisticated kinetic model upon a reliable statistics would be enormous.
Therefore, in this work we preferred to embed the higher-order effects in the transition rates $W$ under the form of an overall scaling factor $f$.
In order to do this, we start from the memory-expansion expression of the center of mass diffusivity $D_{c.m.}$\cite{Ying1998}
\begin{align}
    D_{c.m.}= \frac{1}{2dN\tau}\left(C^{\delta \mathbf{R}}_0 + 2\sum^{\infty}_{t=1} C^{\delta \mathbf{R}}_t\right),
    \label{eqn:MEM}    
\end{align}
where $d$ is the dimensionality, $\tau$ is the chosen time interval, $N$ is the total number of molecules. $C^{\delta \mathbf{R}}_t$ is the center of mass displacement autocorrelation function which reads
\begin{align}
    C^{\delta \mathbf{R}}_t= \langle \delta \mathbf{R}_0 \cdot \delta \mathbf{R}_t \rangle , 
    \label{eq:autocorr}
\end{align}
where $\delta \mathbf{R}_t= \sum^{N}_i \left( \mathbf{r}_t - \mathbf{r}_{t-1} \right)$ is the summation of all molecular displacements between time $t-1$ and time $t$.
Considering a purely Markovian approximation and neglecting all the correlation effects for $t>0$ in Eq. \eqref{eqn:MEM} yields the dynamical mean-field (DMF) expression of center-of-mass diffusivity\cite{Reed1981}, $D^o_{c.m.}$:
\begin{align}
    D^o_{c.m.}= \frac{C^{\delta \mathbf{R}}_0}{2dN\tau}= \frac{\overline{W}a^2}{2dN\tau},
    \label{eqn:DMF}
\end{align}
where $\overline{W}$ is the average jump probability and $a$ is the lattice cell parameter.
The ratio between the infinite-memory diffusivity and the DMF diffusivity, $D_{c.m.}/D^o_{c.m.}$, can be taken as a measure of how memory effects influence the overall diffusion process. 
If such a ratio is below $1$, then the overall effect is a slowing down of diffusion induced by negative correlations in displacements; if overall correlations in displacements are positive, instead, then the ratio $D_{c.m.}/D^o_{c.m.}$ turns out to be larger than 1, this resulting in an increase of diffusivity.
By estimating the correction factor as $f=\left(C^{\delta \mathbf{R}}_0 + 2\sum^{\infty}_{t=1} C^{\delta \mathbf{R}}_t\right) / C^{\delta \mathbf{R}}_0$ and by using Eqs. \eqref{eqn:MEM} and \eqref{eqn:DMF}, we obtain
\begin{align}
    D_{c.m.}= fD^o_{c.m.} = \frac{f\overline{W}a^2}{2dN\tau}.
    \label{eqn:CorrDiff}
\end{align}
Our idea is then to correct the purely Markovian jump rates according to $\overline{W}^{corr}=f\overline{W}$, and then to use such corrected jump rates $\overline{W}^{corr}$ in the numerical CG simulations, rather than $\overline{W}$.
In general, we expect $f$ to be a function of the global density $\langle n \rangle$ and this would cause the evolution operator to depend on a global variable; however, since we want $f$ to be local as well, we can circumvent this problem by replacing the dependence on the global density $\langle n \rangle$ with a \emph{local density guess}, i.e.~a guess of $\langle n \rangle$ on the basis of local occupancies.
More specifically, our choice is to use the average local pair occupancies $\overline{M}_{12}=(n_1 + n_2)/2$ rather than $\langle n \rangle$ as input for the function $f$.
In this way, we easily correct our local operator by embedding the overall effect of time-correlations and yet we keep locality and our approximate DB condition, since the average local pair occupancy $\overline{M}_{12}$ is a conserved quantity during each elementary transition.
Despite its simplicity, this method allows one to estimate the overall correlation effect directly from the analysis of the original MD trajectories, without having to perform further simulations of the reference system \cite{Beerdsen2004}.

\subsection{Numerical simulations}
We performed the FG atomistic simulations by modelling all the atoms involved as Lennard-Jones (LJ) particles. 
The whole methane molecule was represented by a single LJ bead, following the widely accepted united-atom approximation \cite{Dubbeldam2005}. 
The methane-carbon and methane-zeolite LJ interactions were parameterized according to our previous works \cite{Pazzona2018,Pireddu2019}.

\begin{figure}
  \includegraphics[width=3.25 in]{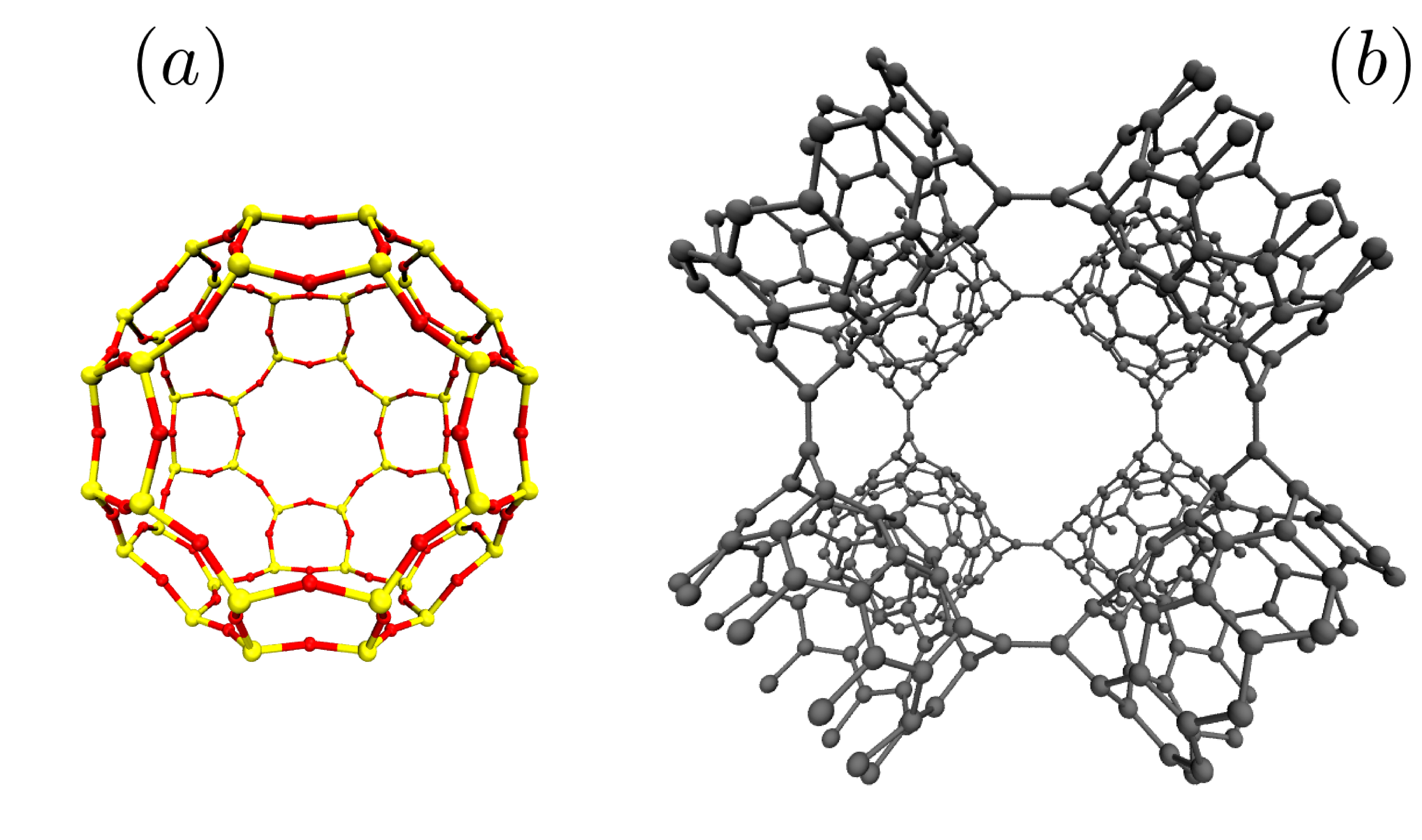}
  \caption{\footnotesize{Atomistic representations of the cages corresponding to the ITQ-29 (subfigure $(a)$) and the ZTC (subfigure $(b)$) materials.}}
  \label{fgr:Cages}
\end{figure}

In all the simulations, the host materials were represented as rigid frameworks.
The ZTC crystalline structure, in its \textit{unrelaxed} version, was downloaded from \url{materialscloud.org}\cite{ZTCcif}, while the ITQ-29 structure was taken from RASPA2's repository on \url{github.com}\cite{ITQ29cif}.
A comparison of the pores of the two host materials is presented in Fig. \ref{fgr:Cages}.

The systems we considered were simulated at the same temperature, i.e.~300 K. 
The reason for this choice is two-fold: we wanted to represent a realistic scenario for room-temperature applications of such systems, and at the same time this temperature was observed to yield a sufficient number of molecular inter-cage jumps in our simulations.

GCMC atomistic simulations were required for the calculation of the IPA parameters, and were performed using an in-house built code with the usual displacement, insertion and deletion trial moves \cite{Frenkel2002}, whereas all MD simulations were performed by using the open-source software LAMMPS.\cite{Plimpton1995}
We computed the MD trajectories for several methane loading values ($\langle n \rangle=$ 1,2,... up to 14 for the ITQ-29, and 15 for the ZTC) in terms of average number of guest molecules per pore, and considering a $3\times3\times3$ cells of the host materials, where every cell contains a single pore.
Our choice for such maximal loading values is motivated by that fact that, in the zeolite case, we did not observe any inter-cage jump $\langle n \rangle > 14$, whereas in the ZTC case, loading values above $\langle n \rangle = 15$ resulted in the emergence of new inter-cage adsorption sites, which would require a much more complicated CG mapping.
Also, since we wanted to highlight the comparison between the two materials, we chose a similar range of conditions for the two systems.
In both cases, we obtained the methane trajectories by assuming periodic boundary conditions (PBCs) within the NVT ensemble; temperature was kept approximatively constant at 300 K through a Nos\'e-Hoover thermostat; every MD simulation started with a $0.5$ ns-long equilibration stage; after equilibration, we sampled the dynamics of the methane-zeolite system for $10$ ns (while saving molecular configurations every 1 ps), and the dynamics of the methane-ZTC system for 1 ns (while saving configurations every 20 fs).

In order to prove the accuracy of our method and to study the collective diffusivity in such two systems, we also performed numerical simulations of the CG models. 
Such simulations were conducted by applying the previously parameterized local operators to the lattice models of the reference systems and sequentially updating the states of connected pairs of cells. 
The evolution algorithm of our lattice models is designed as follows. Each simulation stars with initialization of the starting lattice occupancy configuration $\mathbf{n}$, then for each time-sweep the following scheme is used:
\begin{itemize}
    \item[$(1)$] we randomly extract a pair of connected cells out of all the connected pairs in the CG system (the same pair may be invoked more than once during the same time-sweep); 
    \item[$(2)$] we generate all possible outcomes $\mathbf{m}'$ consistently with the total mass conservation and calculate the rate $W^{corr}(\mathbf{m}' \mid \mathbf{m})$;
    \item[$(3)$] we randomly pick a new state $\mathbf{m}'$ according to the probability distribution $W^{corr}(\cdot \mid \mathbf{m})$, and then update the local occupancies;
    \item[$(4)$] if the number of pairs invoked during the current time-sweep turns out to be equal to to the total number of connected pairs, then the current time-sweep is concluded; otherwise, we return to step $(1)$.
\end{itemize}
We empirically estimated the behaviour of collective diffusivity as a function of the loading by using the Boltzmann-Matano (BM) method. 
This method was first introduced by Matano to study the interdiffusion of different metallic species in the proximity of the intermetallic interface \cite{Matano1933}, but it was also successfully applied to the study of collective diffusion of particles in lattice models \cite{ZK2000,ZK2001}.
The BM analysis is conducted on the time-dependent profile of adsorbate density along a chosen direction, obtained from the spread of a step-like initial profile. 
The spread is numerically simulated according to the lattice CG dynamics.
The relation between density profile and collective diffusion coefficient is the following:
\begin{align}\label{eqn:BM}
    D_c(\langle n \rangle)= \frac{1}{2t}\left( \frac{\partial \rho}{\partial x} \right)^{-1}\int^{\langle n \rangle}_{0} \left( x-x_M \right) dx ,
\end{align}
where $\rho$ is the density, $t$ is the time considered for the spread of the initial profile, $x$ is the chosen direction for the analysis and $x_M$ is the position of the Matano plane, which is chosen to fulfil the condition $\int^{n_{max}}_{0}(x-x_M) dx= 0$, with $n_{max}$ as the maximum occupancy.

The simulations used for the Boltzmann-Matano analysis were conducted with $200\times5\times5$ supercells of the reference materials.
We found this supercell configuration to be the optimal compromise in terms of computational effort and smoothness of density profiles.
We also simulated $3\times3\times3$ supercells of the reference materials in order to compare CG and FG relaxation behaviour in terms of occupancy correlations, and their respective static properties in terms of local occupancy histograms.

\section{Results and discussion}\label{Sec:Res}

\subsection{Jump rates modelling}

\begin{figure}
  \includegraphics[width=3.25 in]{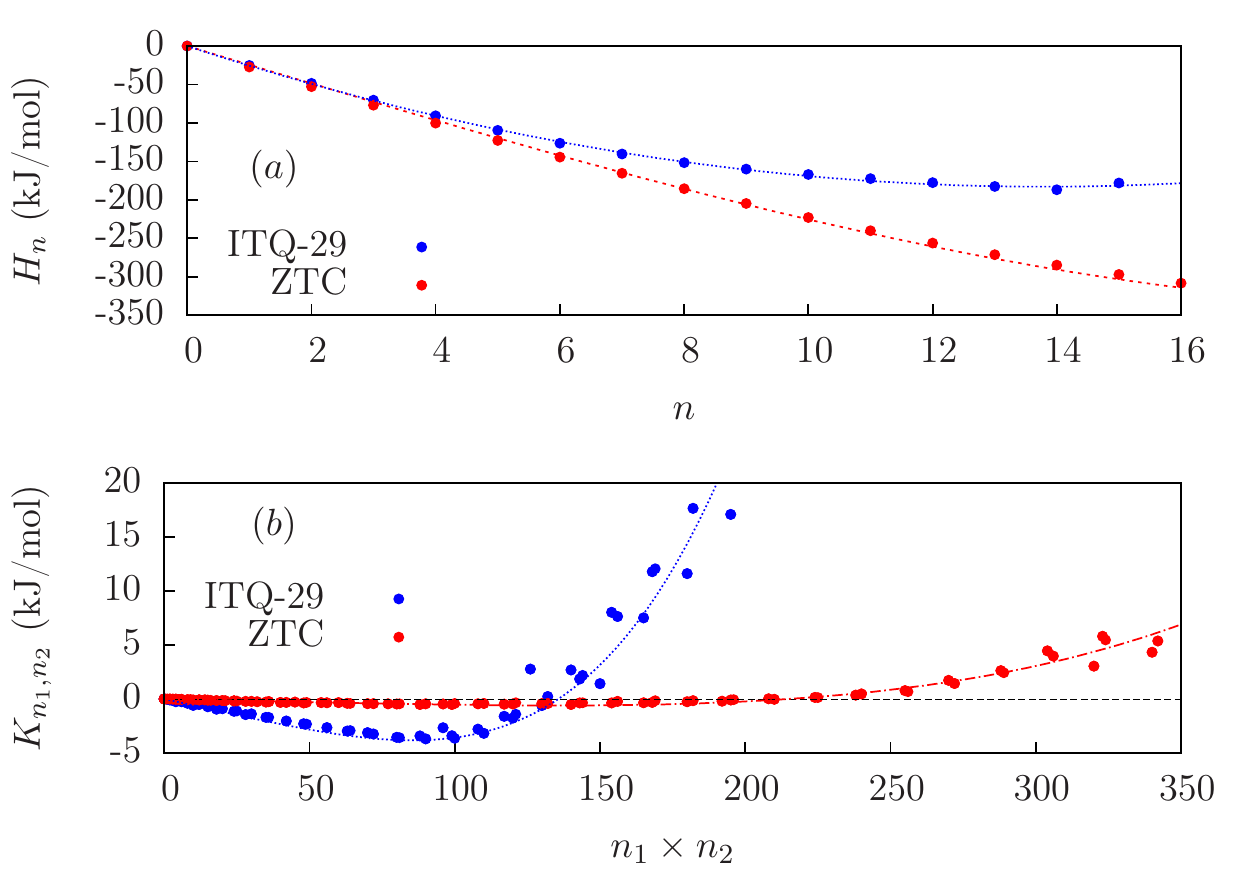}
  \caption{\footnotesize{Free energy parameters in units of kJ/mol as obtained from the IPA CG of the two systems. Subfigure $(a)$ is referred to the single cell contributions $H_n$, while subfigure $(b)$ shows the behaviour of mutual interaction parameters $K_{n_1,n_2}$ as a function of the product of two local occupancies. The points represent the original data, the dashed lines represent the fitted functions used for the CG simulations.}}
  \label{fgr:IPAParams}
\end{figure}

We started our CG procedure by calculating the local free-energy contributions in terms of single-cell $H_n$ and mutual interaction $K_{n1,n2}$ terms for the two systems, within the IPA theoretical framework. 
The results of our free-energy parametrization for the two systems are shown in Fig. \ref{fgr:IPAParams}. 

Our results show that the two systems exhibit a qualitatively similar behaviour in terms of CG thermodynamics. 
The $H_n$ parameters monotonically decrease for the two systems with a progressive trend flattening at high densities.
The mutual interaction parameters show an attractive regime at moderate densities i.e. $n_1 \times n_2 \leq 12$ for the ITQ-29, and $n_1 \times n_2 \leq 15$ for the ZTC system. 
Conversely, for higher values of loading, both systems exhibit a positive and relatively fast-growing mutual interaction contribution.
Such effect reflects an overall repulsion between high-density regions of the host materials.
Despite qualitative similarities between the two systems, for the ZTC case we observe a deeper $H_n$ contribution, indicating that, for a given value of density, the number of favorable configurations in the methane-ZTC system is larger than the ITQ-29 case.
This is a direct consequence of the larger free volume present in the ZTC material, and of the weaker localization of the guest molecules.
In fact, the presence of preferential methane adsorption sites in the ITQ-29 is well known \cite{Demontis1997}, while our results yielded a more uniform distribution of methane positions within the ZTC.
Also, we found weaker mutual interactions in the ZTC system as compared to the ITQ-29.
We believe that this is a consequence of the fact that, in ZTC, spatial correlations between methane molecules localized in neighboring cages are relatively low, due to the weaker confinement effect of the host material.
We assessed the quality of the free-energy parameters by comparing the reference FG occupancy histograms with the ones obtained from CG simulations. We found a satisfactory agreement for both systems; such results are shown in the Supporting Information.

\begin{figure}
  \includegraphics[width=3.25 in]{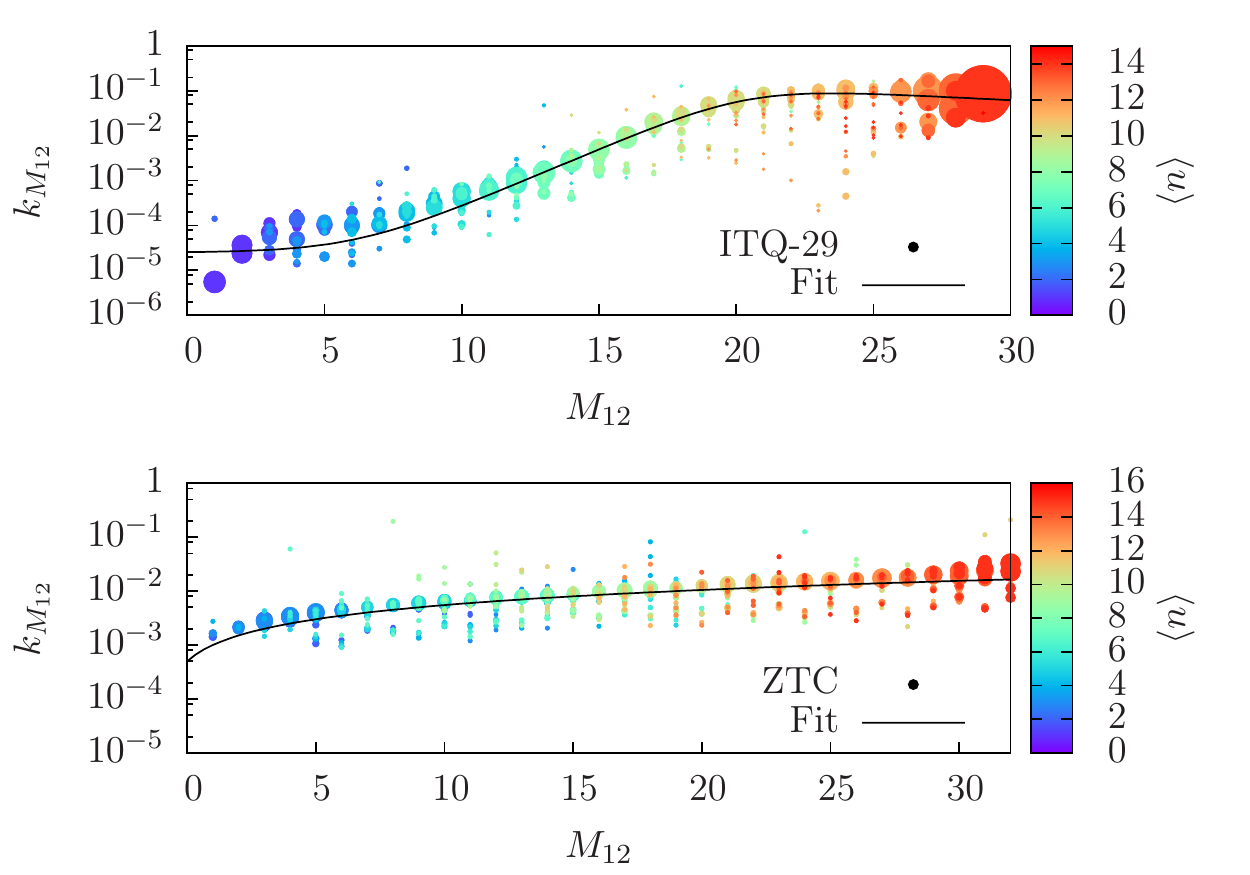}
  \caption{\footnotesize{Fitting of $k_{M_{12}}$ for the ITQ-29 (upper subfigure) and ZTC (lower subfigure) systems. The $y$-axis represents the kinetic prefactor $k_{M_{12}}$, while the $x$-axis represents the summation of the local occupancies $M_{12}=n_1+n_2$. Each point represents a transition observed during the MD simulations, sized according to the probability of the starting configuration and coloured according to the loading of the simulation where such transition occurred. The black solid lines represent the models used in the CG simulations.}}
  \label{fgr:Prefactors}
\end{figure}

In Fig \ref{fgr:Prefactors} we show the fitting of the kinetic prefactor $k_{M_{12}}$ (we remind that $M_{12}$ is the sum of the occupancies of the two cells of the neighboring pair considered during every inter-cell jump event) for the two systems we considered. 

The behaviour of this quantity changes significantly from the zeolite to the ZTC case.
In the first case, we clearly distinguish two regimes: for $M_{12}<20$,  $k_{M_{12}}$ grows relatively fast, following an exponential trend; above $M_{12}=20$, the prefactor mildly decreases. 
Conversely, we found a simpler and more uniform behaviour in the methane-ZTC system. 
In this case, $k_{M_{12}}$ seems to increase linearly respect to the local occupancy.
In order to perform our CG simulations, both data sets were fitted to obtain the $k_{M_{12}}$ function for the two systems;
the fitting models were designed by prioritizing the most frequent events, for which we expect a better accuracy in the jump probability estimation.
Hence, we gave priority to the transitions associated with a larger probability for the initial and the final state.
A more detailed description of the models and parameters used can be found in the Supporting Information.

\subsection{Dynamical correlations}

\begin{figure}
  \includegraphics[width=3.25 in]{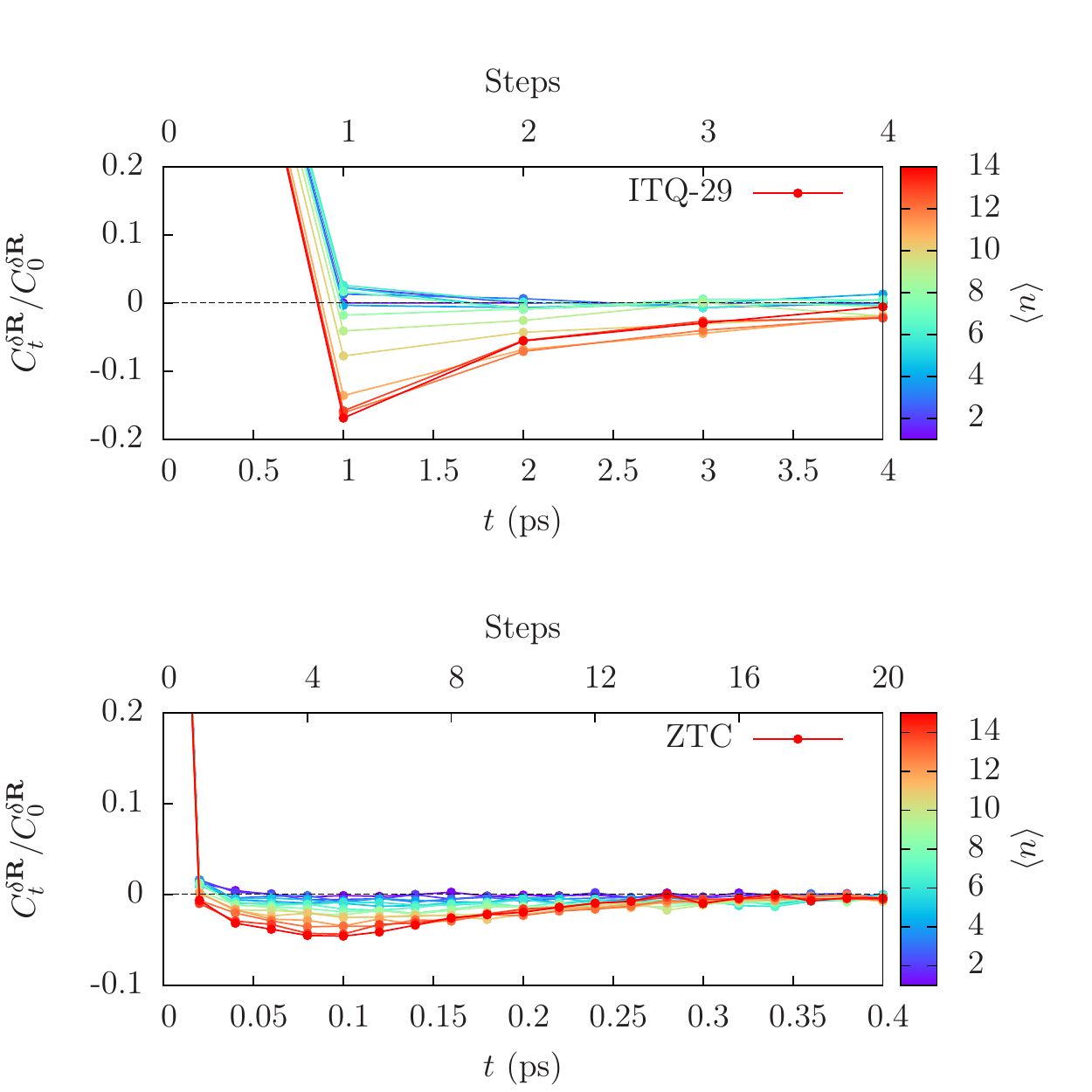}
  \caption{\footnotesize{Normalized center of mass displacement autocorrelation as a function of time, from the MD reference simulations. The upper subfigure is referred to the ITQ-29 system, while the lower subfigure is referred to the ZTC system. The colour represents the loading associated to each MD simulation.}}
  \label{fgr:CMDACF}
\end{figure}

We used the MD trajectories in order to estimate the displacement autocorrelation function $C^{\delta \mathbf{R}}_t= \langle \delta \mathbf{R}_0 \cdot \delta \mathbf{R}_t \rangle$ we previously introduced in the Methods section.
For this calculation we only considered the displacements involving inter-cage jumps, in order to filter-out all the intra-cage dynamical effects.
The results of this analysis are shown in Fig. \ref{fgr:CMDACF}.

For each system, the total number of steps was chosen in such a way as to guarantee a sufficient convergence of the autocorrelation function (Eq.~\ref{eq:autocorr}) to zero.
For the zeolite system, we found that the displacement autocorrelation mostly vanishes after 4 simulation steps, which corresponds to 4 ps, thus indicating that for values of time interval $\tau$ larger than 4 ps, memory effects would not be observed at all.
The results also show that memory effects show up mostly as negative correlations between consecutive displacements; this indicates the importance of the backscattering effect, which is a well-known phenomenon occurring during diffusion through micropores \cite{Demontis2005,Dubbeldam2005}; the depth and persistence of such effects change with the global density of guest molecules.
In fact, we observe that larger backscattering occurs for relatively high gas densities values. This suggests that memory effects depend on the correlations between sorbate molecules.
A similar effect is also observed for the ZTC system, for which we obtained a negative correlation effect that vanishes above 0.4 ps; in this case, memory effects tend to decay faster as compared to the ITQ-29 system, indicating a more efficient thermalization.
However, considering that for this system we chose a time step equal to 0.02 ps (much shorter than the methane-zeolite case), memory effects vanish after 20 consecutive steps; therefore, under the viewpoint of iterations in the CG model, the backscattering effect is more persistent within the ZTC host.

\begin{figure}
  \includegraphics[width=3.25 in]{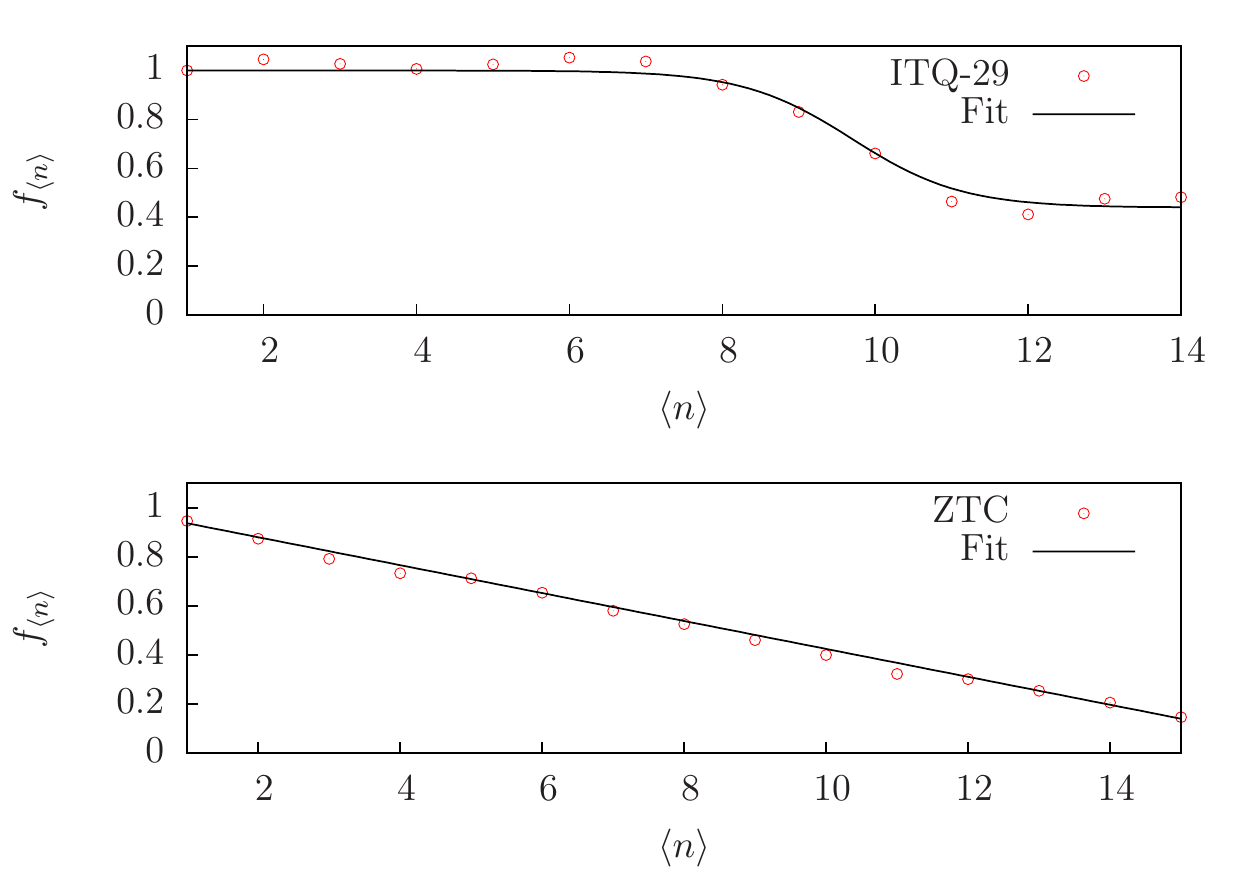}
  \caption{\footnotesize{Correlation factor $f_{\langle n \rangle}$ as a function of the loading. The upper subfigure is referred to the ITQ-29 system, while the lower subfigure is referred to the ZTC system. The results from the MD simulations are represented by red circles, while our fit is represented by a solid black line.}}
  \label{fgr:CorrFact}
\end{figure}

We used the center-of-mass displacement autocorrelation functions to estimate the correlation factor for every loading;
results are shown in Fig. \ref{fgr:CorrFact} for both systems.

In general, we found dynamical correlations to slow down the diffusion process in both the systems we considered.
However, we also found significant differences between the two systems in terms of the behaviour of correlation factors as functions of the loading: for the zeolite system, we observe a sigmoid-like decay for $f$, while for the carbon material we obtained a simple linear decay.
Such differences are due to the presence of different microscopic mechanisms contributing to the decay memory effects and to thermalization; resolving such mechanism would require detailed molecular-level investigations of dynamical correlations, which goes beyond the scope of this work (where we are focusing more on the coarse-graining than on the molecular-level analysis of the reference FG systems) and will be the object of further contributions.
For our purposes, the correlation factor is as a measure of the non-Markovianity of the diffusion process; in fact, $f$ is equal to 1 only if the diffusion is Markovian, which means that memory effects are lost between each time step. 
We observed such condition in the ITQ-29 system at moderate densities ($\langle n \rangle \leq 7$), and in the ZTC system at $\langle n \rangle = 1$. 
Our results suggest that for the systems we considered, a purely jump rates-based modelling of diffusion (i.e., if we kept $f=1$ under all circumstances) would be accurate only for very low sorbate densities; for higher densities, ignoring the dynamical correlations would result in overestimating the jump rates and, consequently, the diffusivity as well. %

\subsection{Diffusivity}

\begin{figure}
  \includegraphics[width=3.25 in]{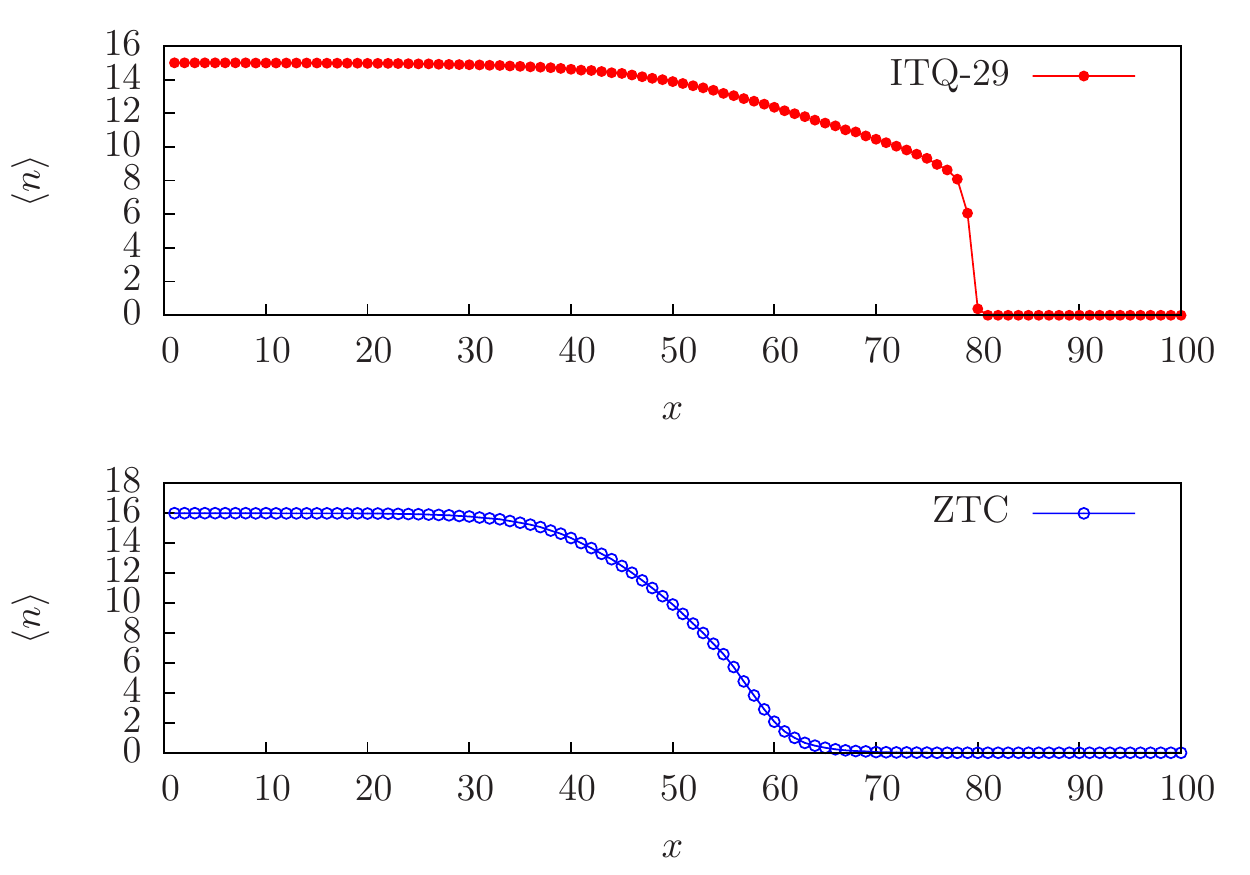}
  \caption{\footnotesize{Density profiles obtained from the Boltzmann-Matano simulations of the two systems. The upper subfigure is referred to the ITQ-29 system, while the lower subfigure is referred to the ZTC system. The plots show only a half of the actual extension of the systems along the $x$-axis. The ITQ-29 profile was obtained by simulating the dynamics for $7$ ns, while in the ZTC case we simulated the system for $0.6$ ns. In both cases, the density profiles were averaged over $100$ replicas of the sytems.}}.
  \label{fgr:Profiles}
\end{figure}

We estimated the collective diffusivity as a function of loading, through the Boltzmann-Matano analysis of CG simulations. 
For both the systems, we simulated the relaxation of the density profiles, according to the procedure described in the Methods section.
In Fig. \ref{fgr:Profiles}, we show a comparison of the density profiles for the two systems. 

For the ITQ-29 system, we observed a first slow decay before a fast step-wise decay of the profile occurring at $\langle n \rangle < 8 $. 
This is the consequence of dramatic differences between diffusivities at low and high densities. 
The sudden decay of the profile is particularly tricky for the numerical BM analysis, because of the lack of points for the lowest densities, an issue that leads to instabilities during the numerical calculation of the diffusion coefficient. 
For this reason, we split the profile relaxation experiment into three separate simulations,each starting with its own initial configurations. 
This procedure is explained in detail in the Supporting Information.
Conversely, the BM profile for the ZTC system is more smooth and qualitatively closer to the shape of the error function, which is related to a concentration-independent diffusion coefficient.\cite{ZK2000}

\begin{figure}
  \includegraphics[width=3.25 in]{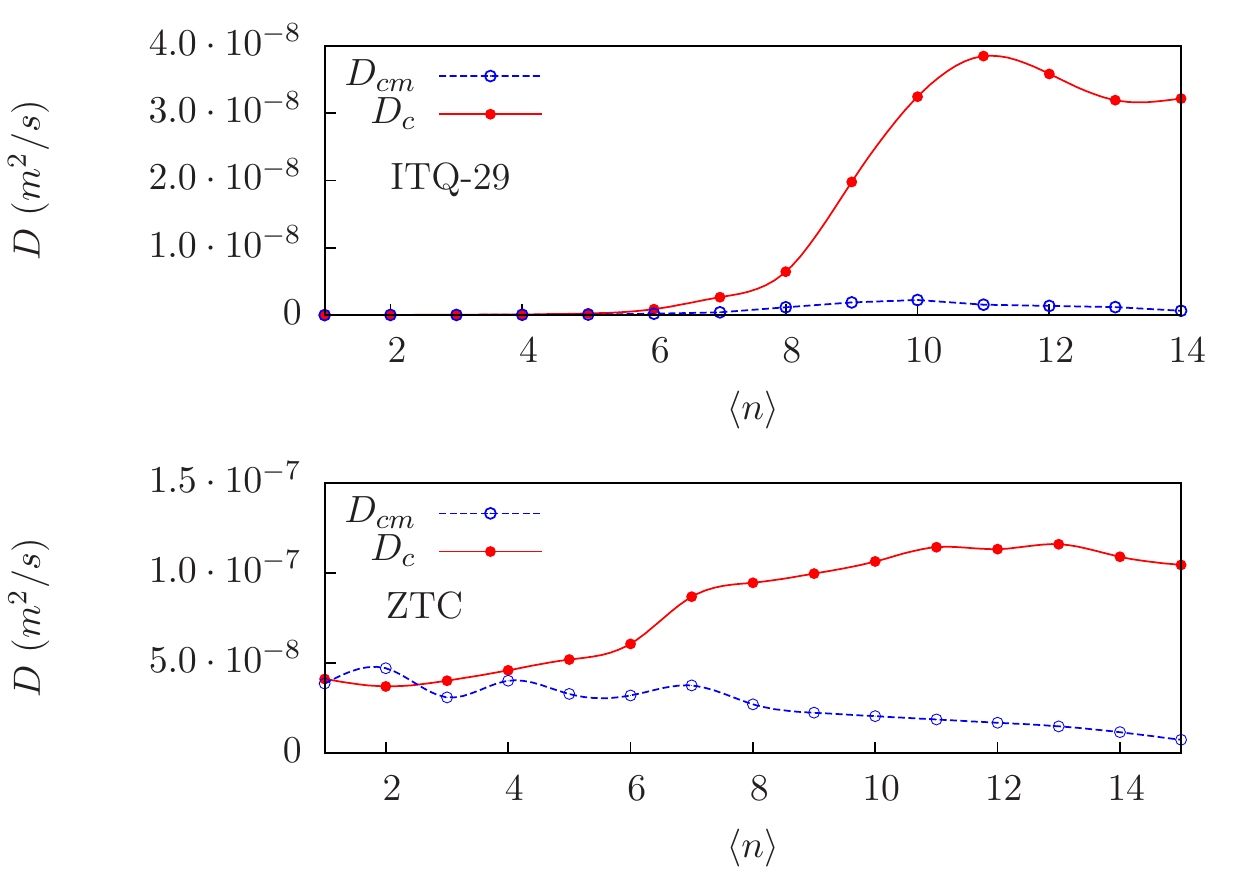}
  \caption{\footnotesize{Center of mass diffusivity $D_{cm}$ (blue points) and collective diffusivity $D_{c}$ (red points) as a function of the loading for both systems. The upper subfigure shows the results for the ITQ-29 system and the lower subfigure is related to the ZTC system. The smooth lines serve only to help the visualization of the results.}}.
  \label{fgr:Diff}
\end{figure}

Our intuitive arguments are confirmed by the trends of collective diffusion coefficients we obtained from numerical calculation.
In order to calculate the diffusivity values for all loadings, we numerically solved Eq.\eqref{eqn:BM} by using the density profiles.
The behaviours of the center of mass diffusivity $D_{cm}$ and the collective diffusivity $D_c$ with respect to the loading are shown , for both systems in Fig. \ref{fgr:Diff}.

The center of mass diffusivity $D_{cm}$ was calculated from the center-of-mass mean-squared displacement, which we estimated from the MD trajectories.
Collective diffusivities were computed, instead, straight from the BM density profiles obtained from CG simulations.
For the ITQ-29 system, we observed a large increase in collective diffusivity for $\langle n \rangle > 7$, with a maximum at $\langle n \rangle = 11$ for which we report $D_c= 3.8 \times 10^{-8}$ m$^2/$s; this corresponds to an increase by a factor of about $10^3$ with respect to the lowest $D_c$ we measured (the lowest $D_c$ was observed at the lowest density investigated, $\langle n \rangle = 1$).
Concerning the behaviour of $D_{cm}$, we found similar results to the ones obtained by Dubbeldam et al., with minor differences due to the slightly different parameterization of the thermostat used in the NVT simulations.\cite{Dubbeldam2005,Beerdsen2006prl}
In our case, we observed a maximum of $D_{cm}= 2.2 \times 10^{-9}$ m$^2/$s for $\langle n \rangle = 10$, which is about $10^2$ times higher respect to the lowest value reported at $\langle n \rangle = 1$.
Results for the ZTC system show milder variations of diffusivities with respect to the loading (we report an increase of collective diffusivity up to $1.2 \times 10^{-7}$ m$^2/$s at $\langle n \rangle = 13$), but larger collective diffusivities for the whole loading range.
This difference with respect to the methane-zeolite case is mainly due to the larger free volume of the material and, in particular, to the larger windows connecting adjacent cages.
Conversely, we obtained a roughly linear decay of the center-of-mass diffusivity with respect to the loading. In fact, at $\langle n \rangle= 16$ it reaches about half of the initial value; we found this behaviour to be surprisingly similar to the one reported by Beerdsen et al.~for the LTL channel-like zeolite,\cite{Beerdsen2006,Beerdsen2006prl} despite ZTC and LTL being very different both in chemical composition and in framework topology, thus suggesting that the diffusive behaviour of methane in ZTC is closer to the diffusion in tube-like structures rather than in cage-like structures like ITQ-29.

\begin{figure}
  \includegraphics[width=3.25 in]{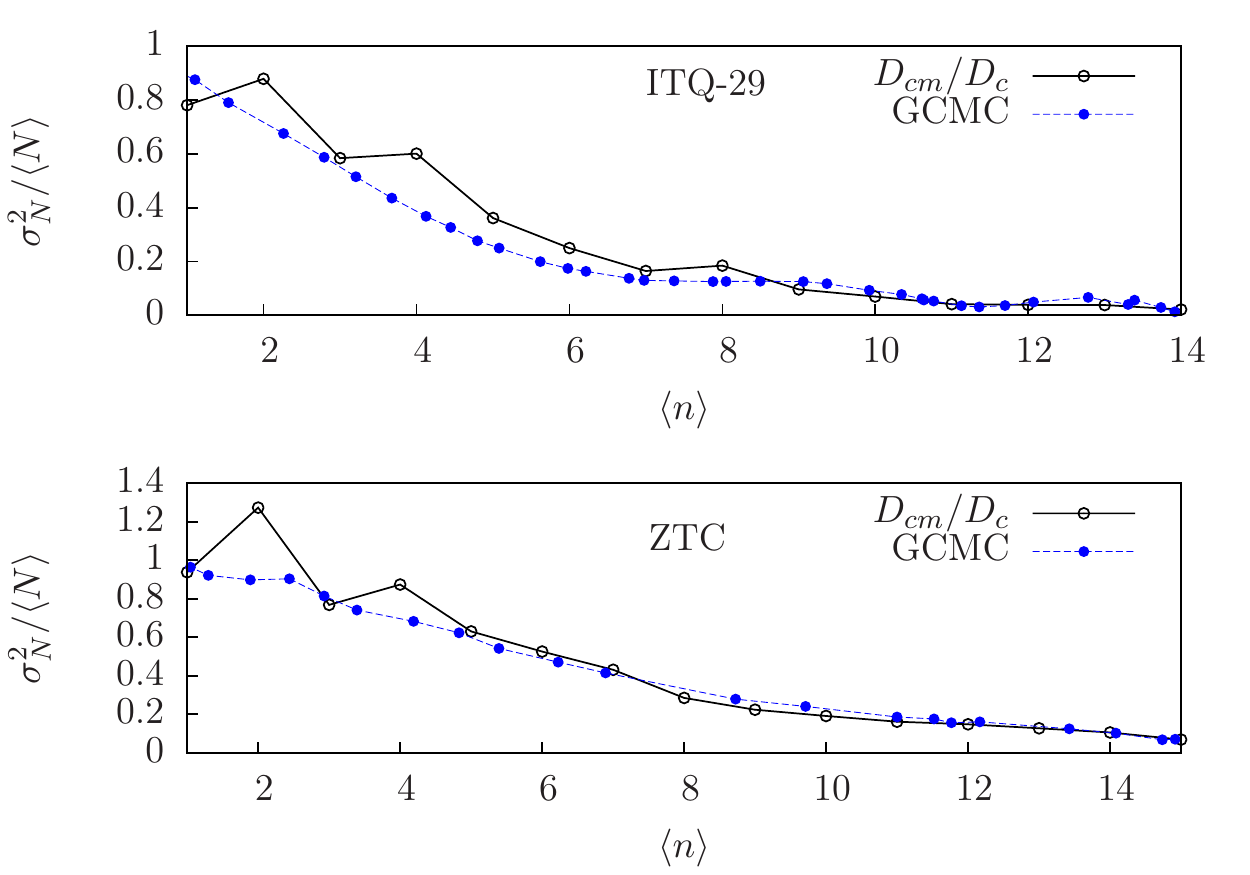}
  \caption{\footnotesize{Comparison between the reduced variance of the total number of particles and the ratio $D_{cm}/D_{c}$ for the two systems. The top subfigure represents the results for the ITQ-29 system, while the bottom subfigure shows the results for the ZTC system. The atomistic GCMC results are shown in blue, while the ratio between diffusivities is shown in black. The center of mass diffusivity $D_{cm}$ is computed from the MD trajectories, while the $D_{c}$ values are obtained via BM analysis.}}
  \label{fgr:ETF}
\end{figure}

The consistency of our diffusivity calculations was validated by comparing the reduced variance $\sigma^2_N/\langle N \rangle$ as computed from the ratio $D_{cm}/D_{c}$, with the same quantity as estimated through GCMC simulations of the FG reference systems. 
The results of this comparisons are shown in Fig. \ref{fgr:ETF}.
We found a satisfactory agreement between the different data sets for both systems, especially at mid-high values of density. The FG/CG data sets for ZTC exhibit a better overlap compared to the results obtained for the ITQ-29 case.
We remark that for the latter system we had a drastically lower number of observed transition in the MD simulations as compared to the ZTC case.
Hence, we believe that better results could be achieved by longer MD simulations of the reference zeolite system, which would yield more accurate and more robust statistics.

\subsection{Decay of occupancy correlations}
\begin{figure}
  \includegraphics[width=3.25 in]{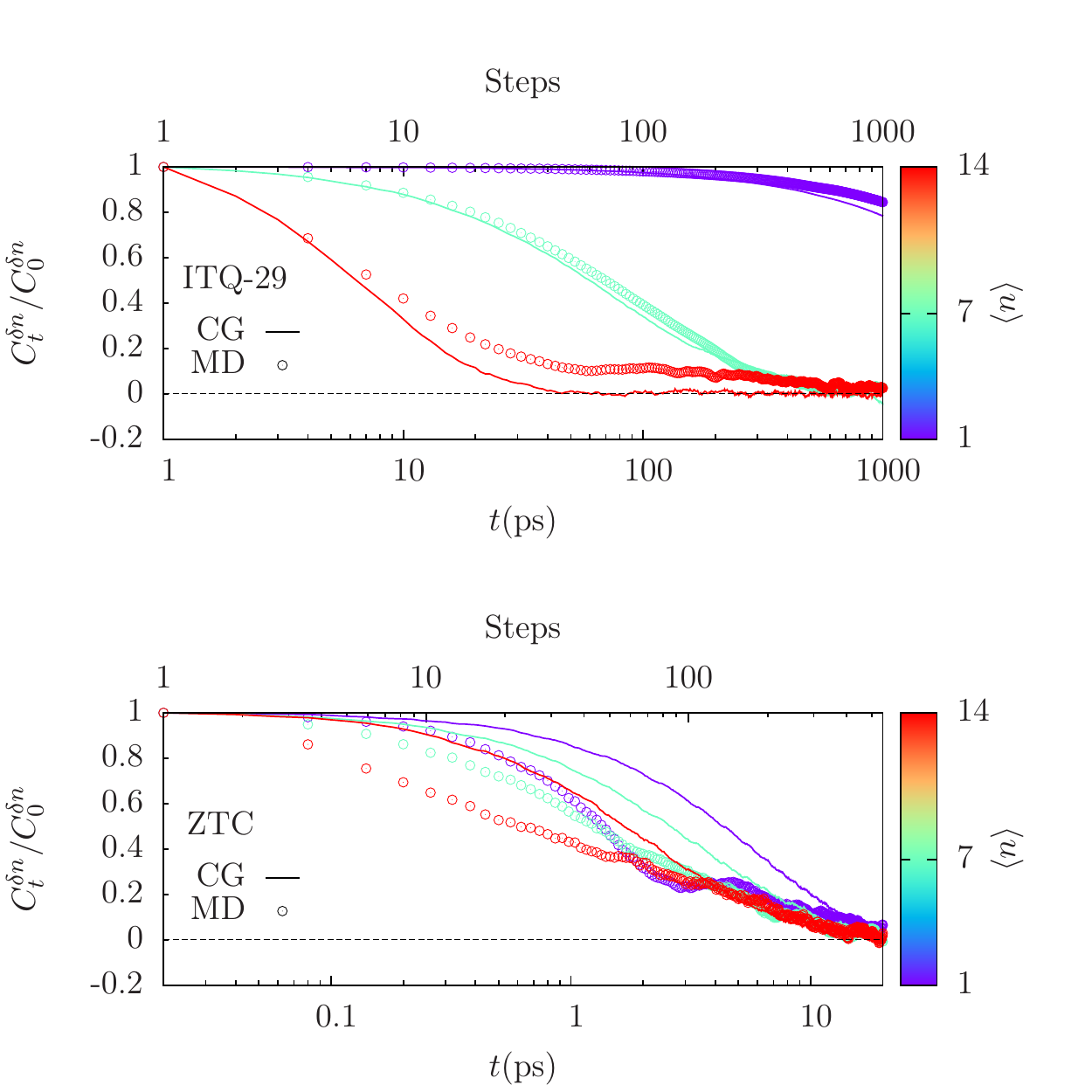}
  \caption{\footnotesize{Normalized occupancy fluctuations autocorrelation functions as a function of time for the two systems at three different loadings: $1$, $7$, $14$. The upper subfigure is referred to the ITQ-29 system, while the lower is referred to the ZTC system. The empty circles represent the results from MD simulations and the solid lines represent the results of CG simulations.}}.
  \label{fgr:OCCACF}
\end{figure}

Since our model makes use of local densities only, and since we assume periodic boundary conditions, the sorbate center-of-mass can not be tracked without introducing ambiguities. 
For this reason, occupancy autocorrelations should be considered as best candidates for measuring the memory decay in CG simulations, rather than correlations in center-of-mass displacements; to this aim we computed the occupancy fluctuations autocorrelation function $C^{\delta n}_t= \langle \delta n_t \cdot \delta n_0 \rangle$, where $\delta n= n - \langle n \rangle$.\cite{Gomer1990}
Fig. \ref{fgr:OCCACF} clearly shows that occupancy autocorrelations vanish more rapidly in the methane-ZTC system, mainly because of the faster mass-exchange dynamics, and that for both systems faster relaxations occur at higher density values. 

This trend is more evident for the ITQ-29 system, for which we found large differences in $D_c$ between low and high-density regimes---in fact, we observed a direct correlation between the relaxation efficiency and $D_c$. This general trend is also reproduced by the CG simulations. 
However, there are evident differences in the agreement between MD and CG results in the two systems:
for the ITQ-29, we obtain a semiquantitative agreement between MD and CG relaxation behaviours, especially for lower loading values, while results for the ZTC system exhibit larger discrepancies. 
We believe that the poorer agreement between the CG and MD data is due to the more markedly non-Markovian nature of mass exchange processes in the carbon material, in relation with the time scale ($\tau$) we chose for such system. 
In fact, for a purely Markovian process the autocorrelations are expected to decay according to an exponential behaviour;\cite{Janke2002} in the present cases, instead, MD results suggest the presence of more complicated relaxation mechanisms, which cause deviations from simple exponential decays. 
Considering that our model is designed as a first-order Markov chain, our best expectation is the obtainment of an exponential approximation of the reference data.
Higher order or multi-time scale transition rate models could allow for the modelling of more complex dynamics and then for a more quantitative matching of occupancy autocorrelation decays; however, as we mentioned while describing the modelling of the transition function $W$, in that case we would have to face the problem of obtaining statistically meaningful data, necessary to implementing a multivariate transition function, from short atomistic simulations. This will be the object of further contributions.

\subsection{Computational speedup}

Simulating the reference systems with our CG models required a considerable less effort in terms of computational resources. We quantified the efficiency gain in terms of the speedup, $S$, defined according to Merrick et al.\cite{Merrick2007}:

\begin{align}
    S= \frac{t_{MD}}{t_{CG}},
\end{align}

where $t_{MD}$ and $t_{CG}$ indicate the time, in units of seconds, required to perform the same simulation with the MD and CG representations, respectively.
To measure the speedups, we simulated 50 ps of the dynamical evolution of $3\times3\times3$ supercells of the reference systems at different densities. We performed our tests on a single CPU core. The results of the speedup calculations are reported in Tab.~\ref{tab:Speedups}.

\begin{table}
\centering
\begin{tabular}{llll}
    \midrule
    {System} & {$S_{\langle n \rangle=1}$} & {$S_{\langle n \rangle=7}$} & {$S_{\langle n \rangle=15}$} \\ \midrule
    ITQ-29 & 90200 & 94400 & 127400 \\ \midrule
    ZTC & 4280 & 5600 & 6863 \\
    \midrule
\end{tabular}
\caption{Speedup values for both of the systems at different loading conditions i.e. $\langle n \rangle=1$,$7$,$15$.}
\label{tab:Speedups}
\end{table}

The results show that the improvement related to the ITQ-29 system is much larger compared to the one related to the ZTC system,
this being due to the different time scales ($\tau$) we considered for the transitions in the two systems:
every CG iteration for the zeolite corresponds to 1 ps of dynamics, while for the carbon material one CG iteration corresponds to 20 fs. 
The consequence is that for the zeolite, only 50 iterations are required to cover the dynamics of the speedup tests; while for the ZTC we need to simulate the system for 2500 iterations.
We also observe that the speedup is loading dependent, due to the fact that the number of degrees of freedom (DoFs) of our CG representations does not depend on the total number of molecules, but only on the number of simulated cavities of the host materials. 
Conversely, the computational effort of MD simulations is proportional to the loading, since the number of DoFs is proportional to the total number of guest molecules.

\section{Conclusions}\label{Sec:Conclusions}

In this work, we demonstrated a successful way to map atomistic simulations of host-guest systems to occupancy-based lattice models. We focused on the problem of gas molecules confined in microporous materials.
In particular, we chose to study methane gas in two different environments: the widely studied pure-silica ITQ-29 zeolite and the LTA-ZTC, a hypothetical carbon material introduced by Braun et al. obtained by the simulated carbon templating of the LTA-zeolite \cite{Braun2018}.

Our method makes use of statistical data of reference systems as drawn from the results of atomistic simulations: GCMC for static properties and MD for dynamical properties. Our lattice models are equipped with a CG potential function, representing the free-energy of the system, which depends only on local occupancies.
The diffusion dynamics is thought of as a composition of several local elementary inter-cage jump events. In our CG representations, we represented such events by employing a strictly local operator, which represents the transition probability associated with each mass-preserving migration event.
We modelled the local operator by taking into account the local change in free-energy associated with each transition and a purely kinetic part, which is related to the frequency of migration attempts, and we also proposed a simple way to correct the jump rates for the backscattering contribution on the basis of the displacements autocorrelations observed in the MD simulations; by this way, we allowed for CG models to take into account the non-Markovian memory effects observed in the reference FG systems, which may significantly influence the diffusion in such environments.

We assessed the accuracy of our method by comparing the CG and atomistic results from different perspectives: (i) by comparing static properties in terms of occupancy histograms; (ii) by comparing dynamical properties in terms of the ratio between the diffusion coefficients $D_{cm}$ and $D_c$, and in terms of the reduced variance $\sigma^2_N/\langle N \rangle$ calculated from GCMC simulations; (iii) by comparing the relaxation behaviours in terms of the decay of autocorrelation of occupancy fluctuations.
The results showed a very satisfactory agreement between atomistic and CG results, except for the occupancy relaxation behaviour in strongly non-Markovian scenarios. 
More sophisticated models would be able to represent such phenomena with better accuracy and will be the object of further contributions; however, we remark that the (very satisfactory) accuracy of the CG model proposed in this work was achieved from small-scale and relatively short atomistic simulations---in fact, obtaining reliable CG representations from short-scale atomistic simulations was the very purpose of our investigation.

Our results showed significant dissimilarities in the properties of the two FG systems we considered, due to the different structure and chemical composition of the two materials. 
In general, the larger free-volume of the ZTC material led to a weaker localization of the guest molecules resulting in faster inter-cage jump dynamics, more efficient collective diffusion, and weaker inter-cage spatial correlations. 
The diffusivity behaviour with respect to the loading showed the presence of a strong cage effect in the ITQ-29 material, this resulting in a large peak in diffusivity for $\langle n \rangle= 10,11$, thus confirming the results shown in previous studies.\cite{Dubbeldam2005}
Conversely, the methane-ZTC system exhibited a mild increase in collective diffusivity and a weak decrease in $D_{cm}$, thus resulting in the absence of any cage effect and suggesting that this system behaves more as a channel-like material.

Finally, the use of our CG lattice models resulted in a strikingly high computational speedup comparing with the computing time required by the original MD simulations, which allowed for simulating several nanoseconds of dynamics, for very large systems constituted by thousands of the reference materials' unit cells, within a few minutes on a general-purpose computer.

In conclusion, we believe with this work to have established a theoretical framework for the representation of adsorption and diffusion in the mesoscale, starting from the atomistic representation of the reference systems. 
Our approach can be used to test the mesoscale behaviour of hypothetical systems in possible applications such as gas storage, separation of gas mixtures and sensors design for gaseous species.

\begin{acknowledgement}

G.P. is thankful to Regione Autonoma della Sardegna for the Ph.D. fellowship provided under the “POR-F.S.E. 2014-2020” program.
F.G.P.~is thankful to Regione Autonoma della Sardegna for the Contract ``Ricercatore a Tempo Determinato'', 
financed through the resources of 
POR --- FSE 2014-2020 --- Asse Prioritario 3 ``Istruzione e Formazione'' ---
Obiettivo tematico: 10, Priorit\`a di investimento: 10ii, Obiettivo specifico: 10,5, 
Azione dell'Accordo di Paternariato 10.5.12 --- C.U.P.~J86C18000270002.

\end{acknowledgement}

\providecommand{\latin}[1]{#1}
\providecommand*\mcitethebibliography{\thebibliography}
\csname @ifundefined\endcsname{endmcitethebibliography}
  {\let\endmcitethebibliography\endthebibliography}{}


\pagebreak

\begin{center}
{\Huge \textbf{Supporting Information}\par}

\vspace{2 cm}

\end{center}{}

\section{Fitted models for the coarse-grained simulations}
In this section, we report the fitted models used for our coarse-grained (CG) simulations. The fitted models are introduced to represent the local free-energy parameters $H_n$ and $K_{n_1,n_2}$, for the kinetic prefactors $k_{M_{12}}$ and for the dynamical correlations correction factors $f_{\langle n \rangle}$.   

\subsection{ITQ-29}

\begin{align}
    H_n= a_1 n + b_1 n^2, 
\end{align}
with the optimal parameter values $a_1= -26.5$ kJ/mol and $b_1= 0.96$ kJ/mol.

\begin{align}
    K_{n_1,n_2}= n_1 n_2 \left[ a_2 + (n_1 n_2)^3 b_2 \right],
\end{align}
with the optimal parameter values $a_2= -0.06$ kJ/mol and $b_2= 2.4 \times 10^{-8}$ kJ/mol.

\begin{align}
    k_{M_{12}}= \left[ \frac{a_3}{e^{b_3 M_{12}}} + \frac{c_3}{e^{d_3 M_{12}}} \right]^{-1} + e_3,
\end{align}
with the optimal parameter values $a_3=1.915$, $b_3=-0.071$, $c_3=1.603 \times 10^6$, $d_3= 0.600$ and $e_3= 2.5 \times 10^{-5}$.

\begin{align}
    f_{\langle n \rangle}= \frac{a_4}{1 + e^{b_4(\langle n \rangle + c_4)}} + 1,
\end{align}
with the optimal parameter values $a_4=-0.56$, $b_4=-1.4$ and $c_4=9.7$. In the numerical simulations the density $\langle n \rangle$ is estimated on the basis of the local pair occupancies as $M_{12}/2$.

\subsection{ZTC}

\begin{align}
    H_n= a_1 n + b_1 n^3, 
\end{align}
with the optimal parameter values $a_1=-24.4$ kJ/mol and $b_1=0.018$ kJ/mol.

\begin{align}
    K_{n_1,n_2}= n_1 n_2 \left[ a_2 + (n_1 n_2)^3 b_2  \right],
\end{align}
with the optimal parameter values $a_2= -0.006$ kJ/mol and $b_2= 6.0 \times 10^{-10}$ kJ/mol.

\begin{align}
    k_{M_{12}}= a_3 M_{12} + b_3,
\end{align}
with the optimal parameter values $a_3= 5.0 \times 10^{-4}$ and $b_3= 5.0 \times 10^{-4}$.

\begin{align}
    f_{\langle n \rangle}= a_4 \langle n \rangle + b_4,
\end{align}
with the optimal parameter values $a_4=0.057$ and $b_4=0.994$. In the numerical simulations the density $\langle n \rangle$ is estimated on the basis of the local pair occupancies as $M_{12}/2$.

\section{Static properties}
In this section, we show the comparison of the static properties of the two systems in the molecular dynamics (MD) and CG representations, at $300$ K and different loading conditions i.e. $\langle n \rangle= 4,7,10$, representing low-, mid- and high-density regimes.
The selected properties are the following:
\begin{itemize}
    \item $P(n)$, the probability of observing $n$ particles in a single cell of the system;
    \item $P(n_1 + n_2)$, the probability of observing a summation of occupancies $n_1 + n_2$ within a pair of connected cells; 
    \item $P(n_1 \times n_2)$, the probability of observing a product of occupancies $n_1 \times n_2$ within a pair of connected cells.
\end{itemize}
The last two distributions are meant to ease the comparison of the CG and MD data sets for the bivariate distributions $P(n_1,n_2)$ which, in principle, would require a comparison between different surfaces in a 3D space.
\begin{figure}
  \includegraphics[width=3.25 in]{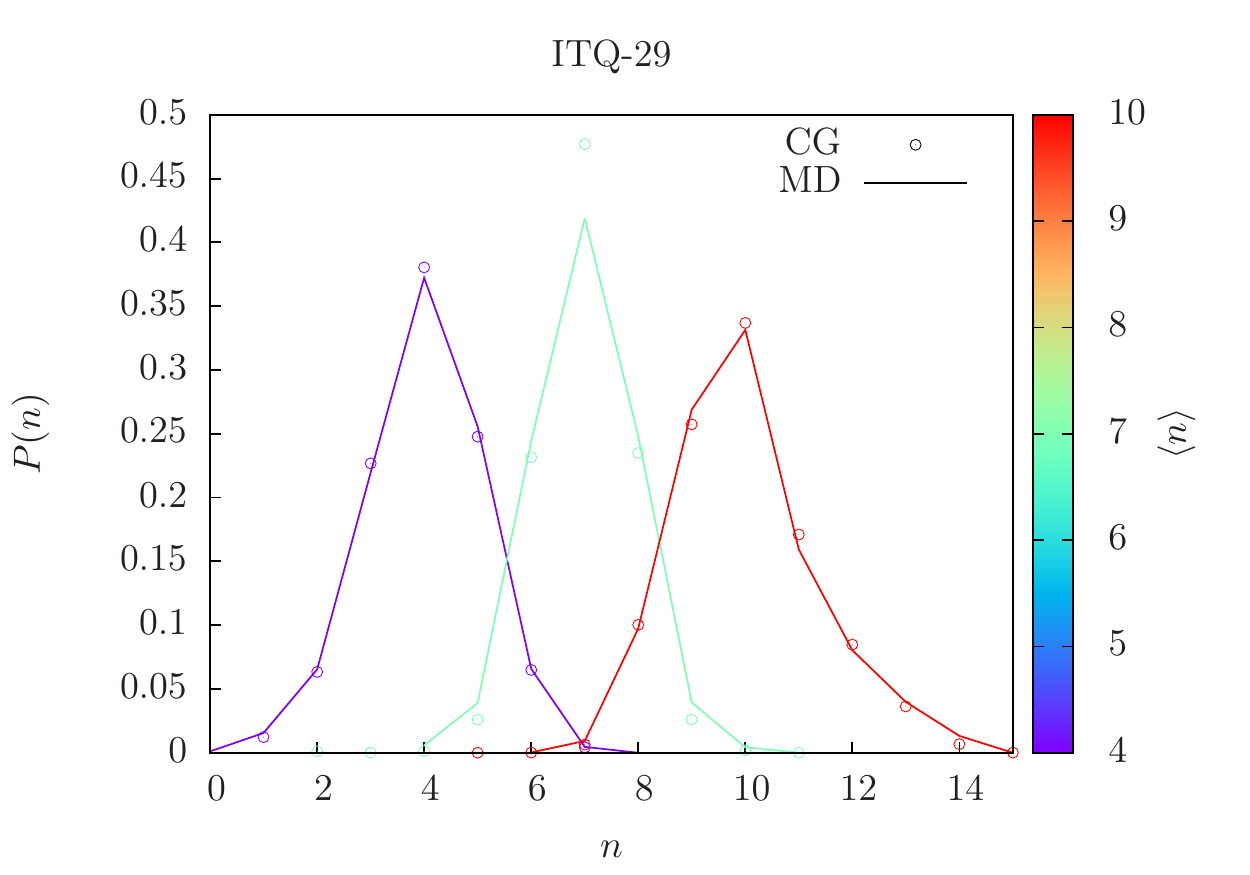}
  \caption{\footnotesize{Single-cell occupancy probability $P(n)$, at different loadings ($\langle n \rangle=$ 4, 7, 10), for the ITQ-29 system. The results from the CG simulations are indicated as empty circles, while the results from MD are represented as solid lines.}}.
\end{figure}
\begin{figure}
  \includegraphics[width=3.25 in]{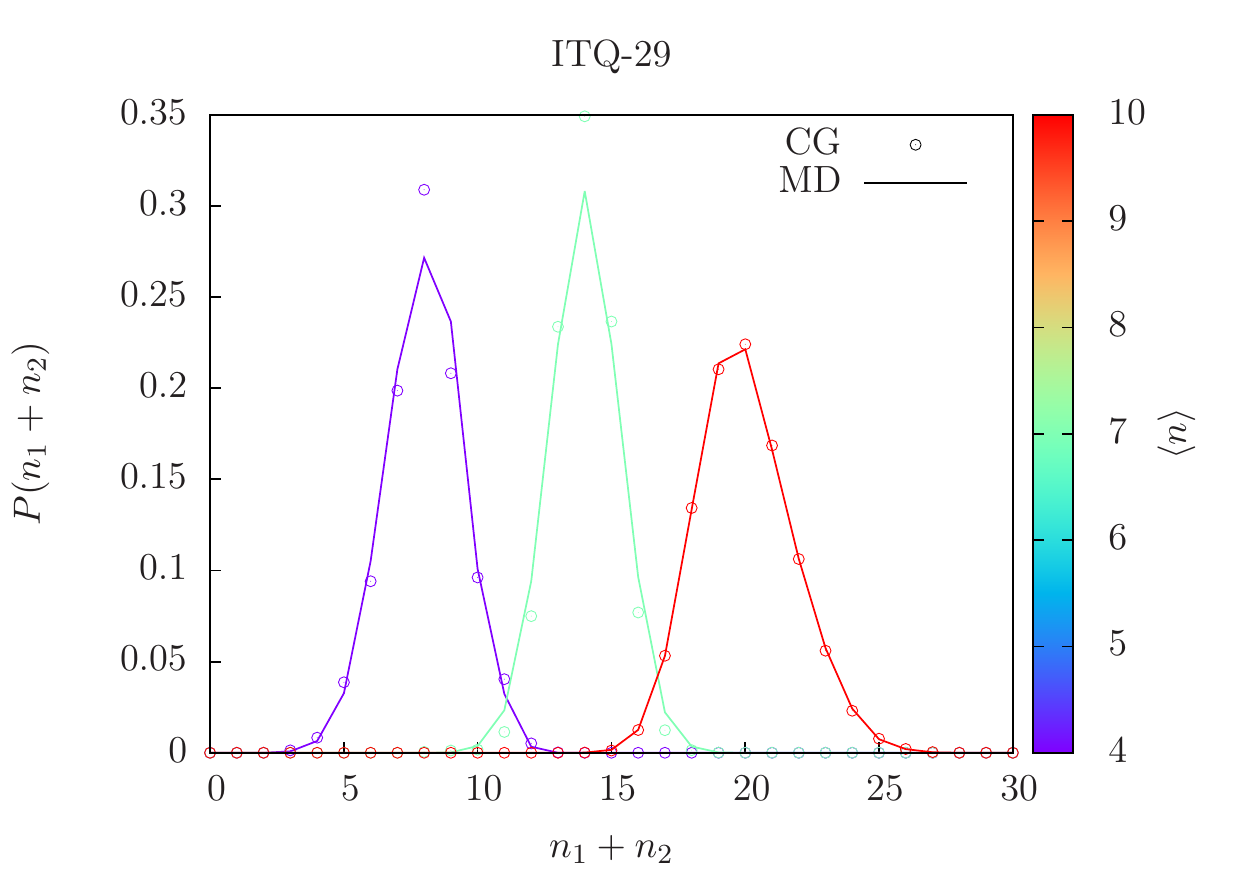}
  \caption{\footnotesize{Neighboring occupancies summation probability $P(n_1 + n_2)$, at different loadings ($\langle n \rangle=$ 4, 7, 10), for the ITQ-29 system. The results from the CG simulations are indicated as empty circles, while the results from MD are represented as solid lines.}}.
\end{figure}
\begin{figure}
  \includegraphics[width=3.25 in]{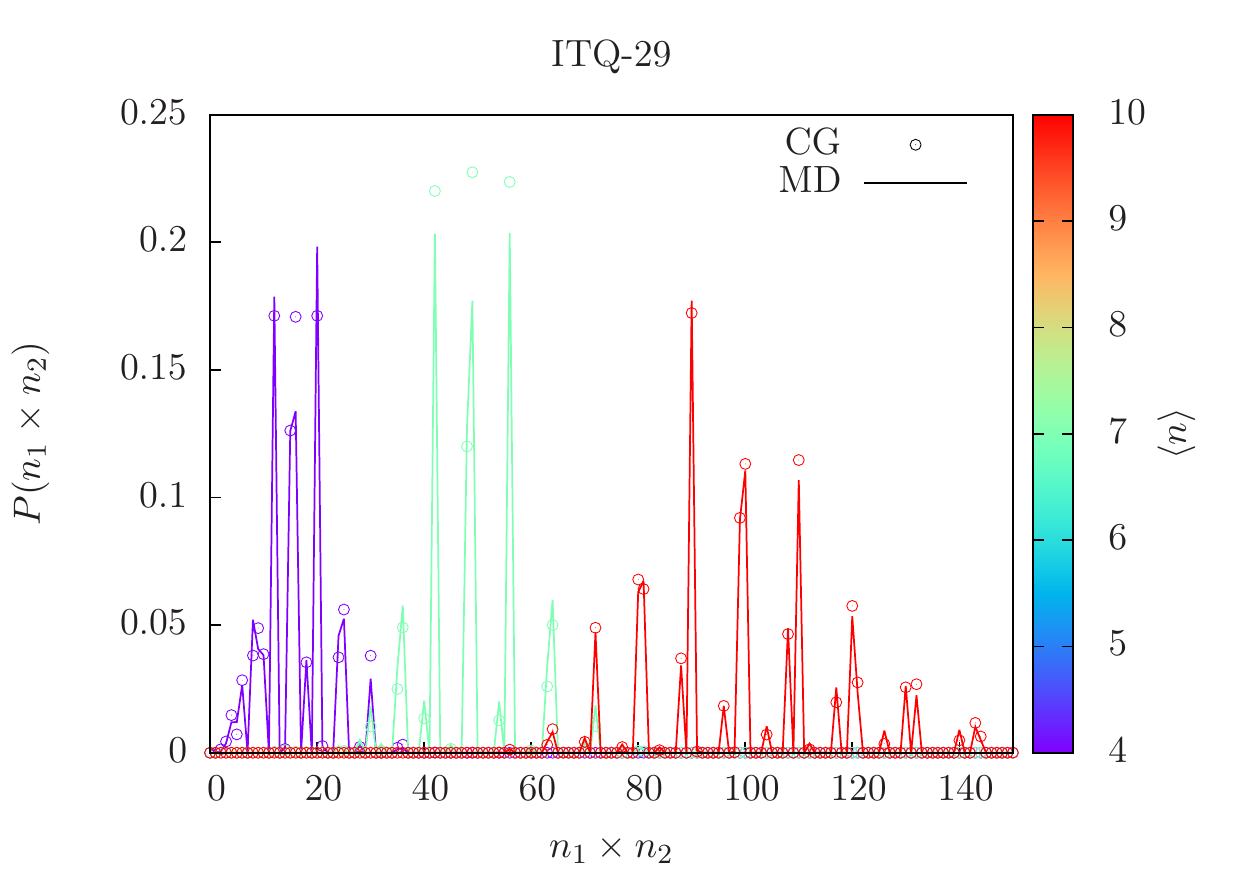}
  \caption{\footnotesize{Neighboring occupancies product probability $P(n_1 \times n_2)$, at different loadings ($\langle n \rangle=$ 4, 7, 10), for the ITQ-29 system. The results from the CG simulations are indicated as empty circles, while the results from MD are represented as solid lines.}}.
\end{figure}
\begin{figure}
  \includegraphics[width=3.25 in]{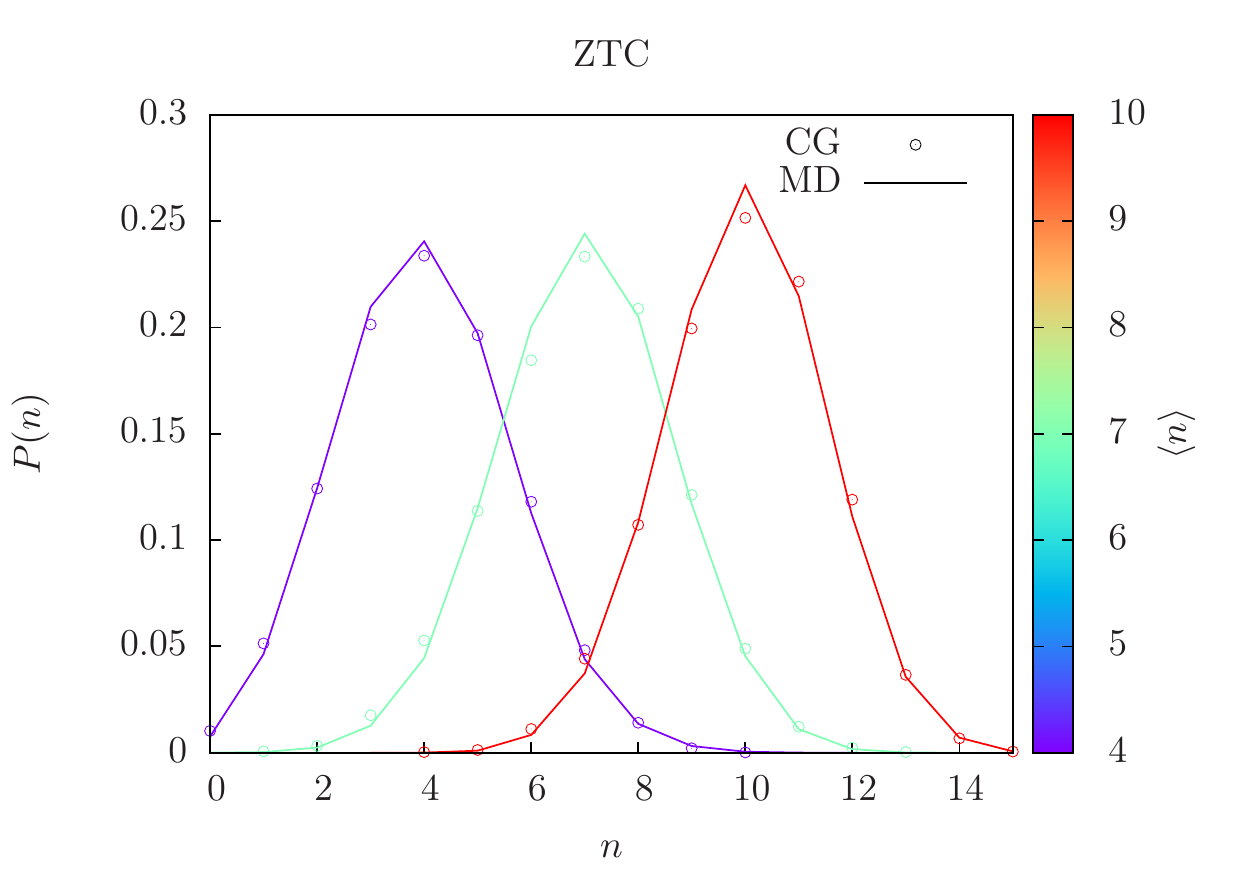}
  \caption{\footnotesize{Single-cell occupancy probability $P(n)$, at different loadings ($\langle n \rangle=$ 4, 7, 10), for the ZTC system. The results from the CG simulations are indicated as empty circles, while the results from MD are represented as solid lines.}}.
\end{figure}
\begin{figure}
  \includegraphics[width=3.25 in]{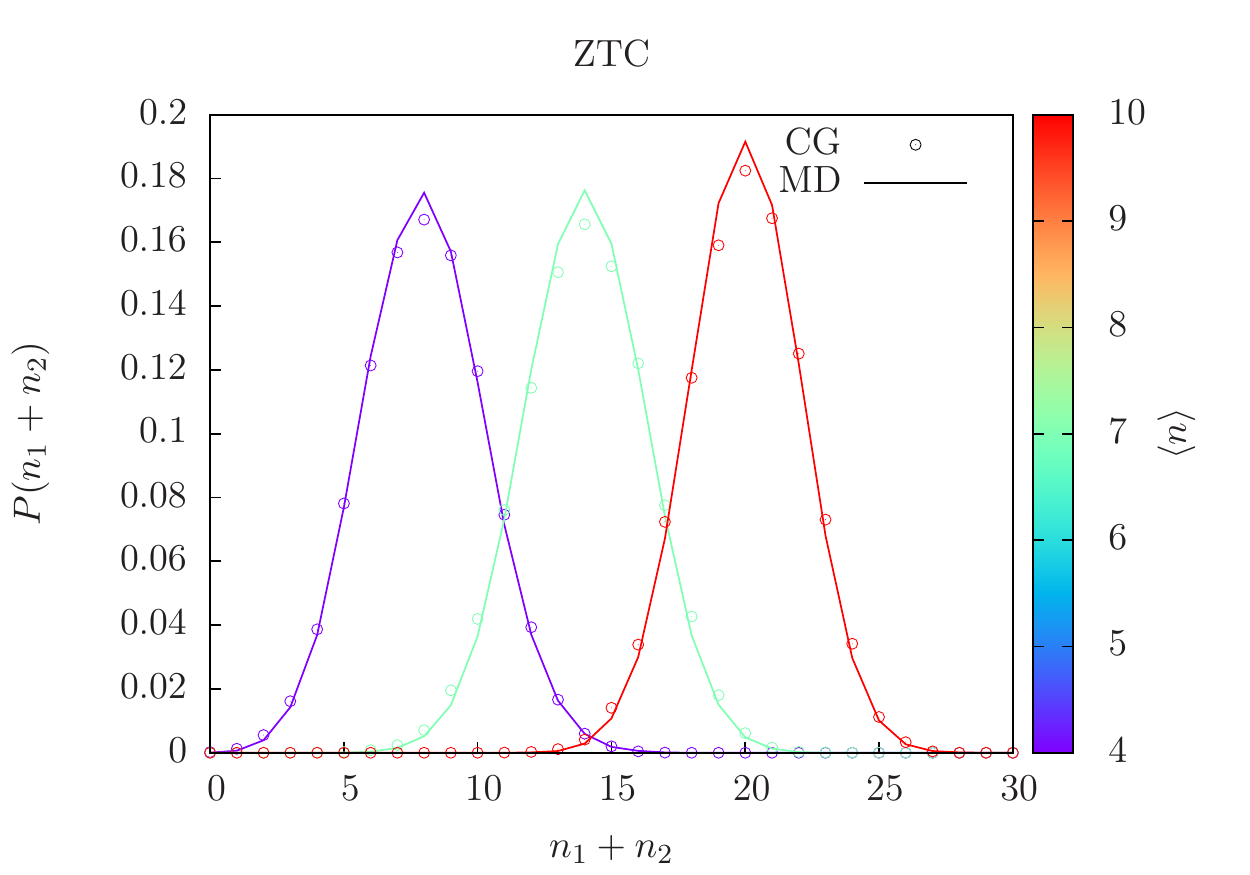}
  \caption{\footnotesize{Neighboring occupancies summation probability $P(n_1 + n_2)$, at different loadings ($\langle n \rangle=$ 4, 7, 10), for the ZTC system. The results from the CG simulations are indicated as empty circles, while the results from MD are represented as solid lines.}}.
\end{figure}
\begin{figure}
  \includegraphics[width=3.25 in]{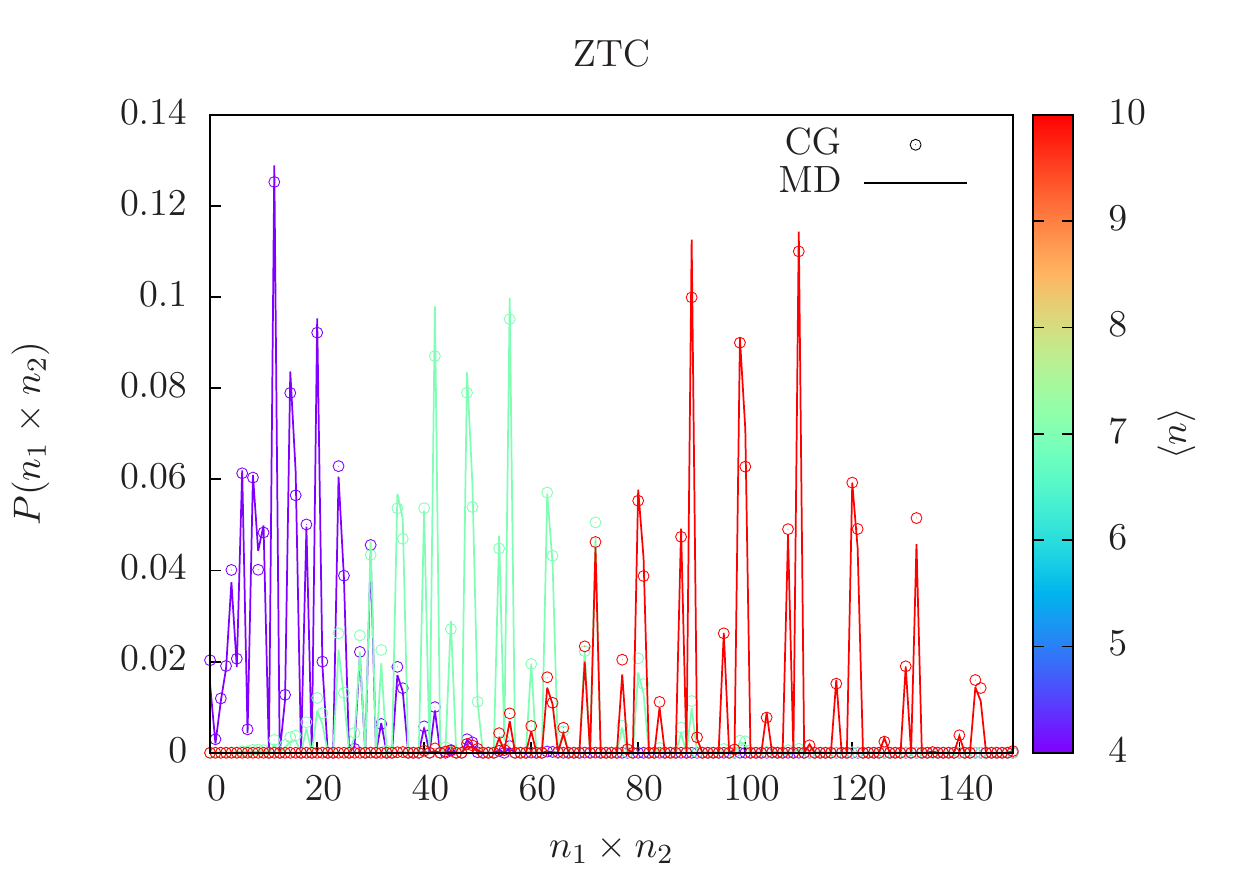}
  \caption{\footnotesize{Neighboring occupancies product probability $P(n_1 \times n_2)$, at different loadings ($\langle n \rangle=$ 4, 7, 10), for the ZTC system. The results from the CG simulations are indicated as empty circles, while the results from MD are represented as solid lines.}}.
\end{figure}

\section{Boltzmann-Matano simulations of the ITQ-29 system}
\begin{figure}
  \includegraphics[width=3.25 in]{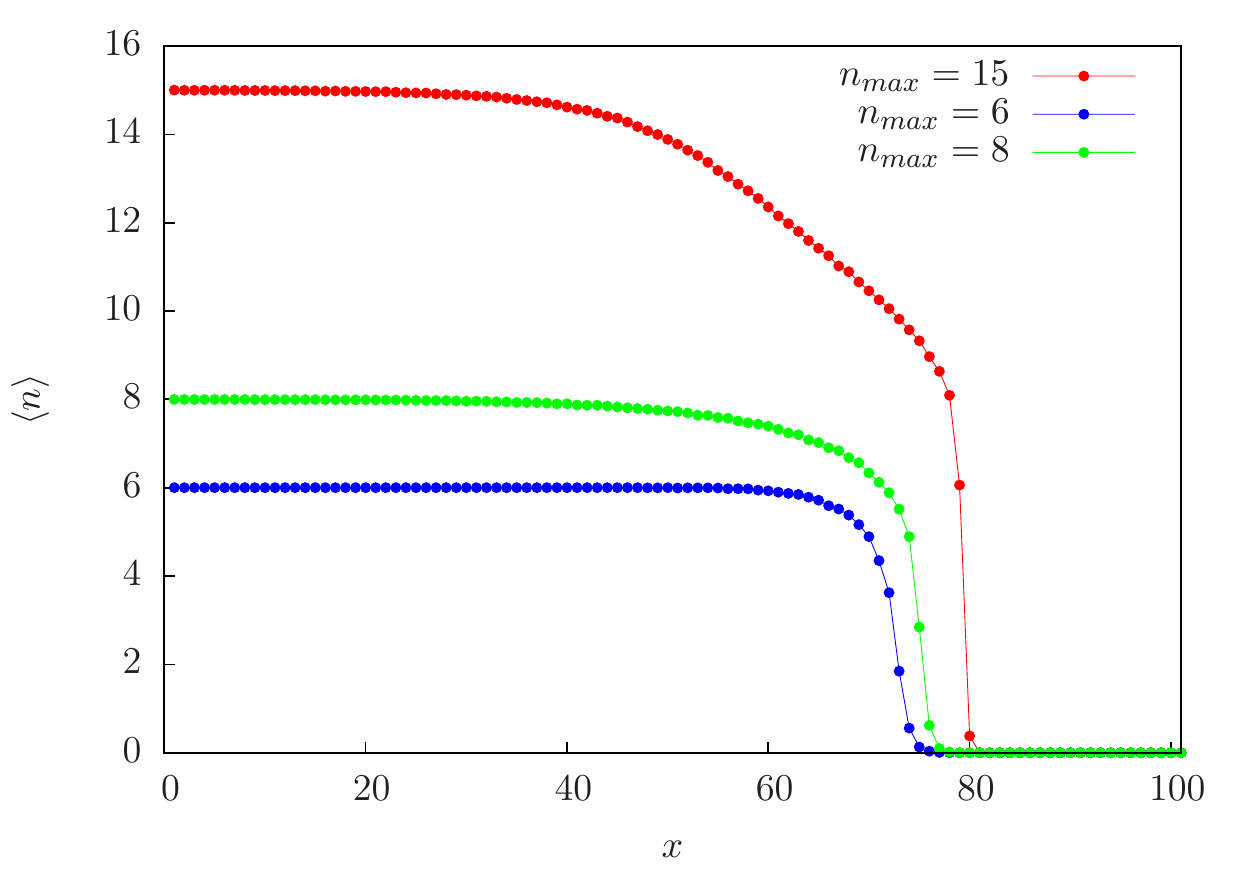}
  \caption{\footnotesize{Density profiles from three different BM simulations of the ITQ-29 system: $n_{max}=15$ is shown in red, $n_{max}=8$ is shown in green, $n_{max}=6$ is shown in blue.}}.
  \label{fgr:ITQ29BM}
\end{figure}

The great changes in collective diffusivity between low- and high-density regimes for this system caused the density profile to be steeply decreasing for $\rho \leq 8$.
This resulted in instabilities during the numerical integration and differentiation of the density profiles.
For this reason, the Boltzmann-Matano (BM) simulations of the ITQ-29 system were conducted in three different versions, each one with a different value for the maximum occupancy $n_{max}$. 
The density profiles we obtained are shown in Fig.\ref{fgr:ITQ29BM}.

We empirically found that setting $n_{max}$ to $15$, $8$ and $6$ was a good compromise between stability and computational effort.
The values of the collective diffusivity $D_c$ for the ITQ-29 system were drawn from the different profiles in order to maximize the stability:

\begin{itemize}
    \item the $n_{max}=15$ profile was used to calculate $D_c$ for $\langle n \rangle \geq 8$;
    \item the $n_{max}=8$ profile was used to calculate $D_c$ for $\langle n \rangle = 7$,$6$;
    \item the $n_{max}=6$ profile was used to calculate $D_c$ for $\langle n \rangle \le 6$.
\end{itemize}

Due to the significantly lower diffusivity respect to high-density regimes, the simulations with $n_{max}=6$ and $n_{max}=8$ required a larger number of simulated iterations ($5$ times respect to the $n_{max}=15$) to reach the profiles shown in Fig. \ref{fgr:ITQ29BM}.



\end{document}